\documentclass[authoryear,preprint,12pt]{elsarticle}

\usepackage{aas_macros}
\usepackage{graphicx}
\usepackage{amssymb}
\usepackage{amsmath}

\usepackage{hyperref}
\usepackage{color}

\renewcommand{\'}[1]{\accent19\ifx#1i\i\else #1\fi}
\newcommand{\sub}[2]{{#1}_{_{\scriptscriptstyle{#2}}}}

\begin{document}
\begin{frontmatter}

\title{A simple model for the location of Saturn's F ring}

\author[lb]{Luis~Benet\corref{cor1}}
\ead{benet@fis.unam.mx}
\author[aj]{\`Angel~Jorba}
\ead{angel@maia.ub.es}

\cortext[cor1]{Corresponding author}

\address[lb]{Instituto de Ciencias F\'{\i}sicas,
Universidad Nacional Aut\'onoma de M\'exico (UNAM)\\
Apdo. Postal 48--3, 62251 Cuernavaca, Mexico}
\address[aj]{Departament de Matem\`atica Aplicada i An\`alisi,
Universitat de Barcelona\\
Gran Via 585, 08007 Barcelona, Spain}


\begin{abstract}
In this paper, we introduce a simplified model to understand the location of
Saturn's F ring. The model is a planar restricted five-body problem defined
by the gravitational field of Saturn, including its second zonal harmonic $J_2$,
the shepherd moons Prometheus and Pandora, and Titan. We compute accurate
long-time numerical integrations of (about 2.5 million)
non-interacting test-particles initially
located in the region between the orbits of Prometheus and Pandora, and address
whether they escape or remain trapped in this region. We obtain a wide region
of initial conditions of the test particles that remain confined.
We consider a dynamical stability indicator for the test particles' motion
defined by computing the ratio of the standard deviation over the average
value of relevant dynamical quantities, in particular, for the mean-motion and
the semi-major axis. This indicator separates clearly a subset of trapped
initial conditions that appear as very localized stripes in the
initial semi-major axis and eccentricity space for the most stable orbits.
Retaining only these test particles, we obtain a narrow eccentric ring which
displays sharp edges and collective alignment. The semi-major axis
of the accumulation stripes of the stable ring-particles can be associated
with resonances, mostly involving Prometheus' outer Lindblad
and co-rotation resonances, but not exclusively. Pandora's inner Lindblad
and co-rotation resonances as well as low-order three-body resonances
typically coincide with gaps, i.e., regions of instabilities.
Comparison of our results with the nominal data for the F ring shows
some correspondence.\\
\end{abstract}


\end{frontmatter}


\section{Introduction}
\label{Sec:Intr}

Saturn's F ring is a fascinating narrow ring with a rich time-varying structure,
which has puzzled dynamical astronomers since its discovery by the Pioneer 11 team
in 1979~\citep{GehrelsEtAl80}. It is located outside Saturn's A ring, close but
beyond Roche's limit for ice, and is believed to be the result of the ongoing
action of competing accretion and disruptive
processes~\citep{BarbaraEsposito2002}. The F ring is narrow, non-circular
and inclined, has azimuthal dependent properties which may change on time, and
displays certain localized radial structures~\citep{PorcoEtAl2005}. It consists
of a dense core ($1-40$~km) embedded in a broader belt of dust
($\sim 700$~km), with additional separated
dusty components named strands, the most prominent form one-arm kinematic
spirals on either side of the core~\citep{CharnozEtAl2005}. It also contains
an underlying belt of forming moonlets~\citep{CuzziBurns1988} that produce
the observed ``fan'' structures~\citep{MurrayEtAl2008,BeurleEtAl2010}, and whose
collisions with the core manifest as jets~\citep{MurrayEtAl2008}. For recent
reviews see~\citet[][sect.~13.5]{ColwellEtAl2009} and \citet{CharnozEtAl2009}.

The F ring is perturbed by the shepherd moons Prometheus and Pandora, that
orbit on either side of the ring. The discovery of these
moons~\citep{SmithEtAl1981,SmithEtAl1982} was initially interpreted as
the success of the confinement produced by the shepherding mechanism,
proposed by~\citet{GoldreichTremaine79}. Yet, the
torques from the shepherd moons do not balance out at the location of the
ring~\citep{ShowalterBurns1982}.
Moreover, the shepherd moons orbit around Saturn on seemingly chaotic
orbits~\citep{PouletSicardi2001,FrenchEtAl2003,GoldreichRappaport2003a,
GoldreichRappaport2003b}. While it is clear that Prometheus and Pandora play an
important role on the dynamics and structure of the
ring~\citep{ShowalterBurns1982,WinterEtAl2007}, it is not clear what is the
actual mechanism that keeps the ring confined at its
location~\citep{Esposito2006}.

Some of the structural phenomena described above have been analyzed previously
through numerical simulations. For instance, using independent test particle
models with periodic boundary conditions along the azimuthal direction,
\citet{GiuliattiWinterEtAl2000} integrated the
equations of motions of the circular restricted three-body problem of Saturn and
Prometheus up to a few tens of revolutions. They showed that, after a close approach
with this moon, ring particles initially located at the strands of the F ring were
scattered inwards and outwards, forming gaps and waves; these were later
confirmed by Cassini, and were named channels~\citep{PorcoEtAl2005}. These
calculations were taken further~\citep{MurrayEtAl2005},
concluding that streamers and channels are part of the same phenomenon and
can be understood in terms of the gravitational interaction with this moon
and its eccentric motion. Other calculations considered the effect of
Pandora during 160 yr, and concluded that the motion of embedded moonlets in
the the ring is likely chaotic, removing it from the F ring
region~\citep{GiuliattiWinterEtAl2006}. \citet{CharnozEtAl2005}
discovered a kinematic spiral strand and interpreted it, based on numerical
simulations spanning 2000 orbital periods, as the effect of interactions
with small satellites in the F ring region. More complex integrations,
including 14 massive Saturn moons and spanning tens of thousands of
Prometheus periods, were conducted by \citet{CuzziEtAl2014}, and led them
to the conclusion that certain regions of stability arise because the
perturbations induced by an encounter with Prometheus are counter-balanced
by subsequent encounters with the same shepherd.

While these investigations have clarified the influence of the shepherd moons in
the variations on the structure of the F ring, the question of its confinement
remains unanswered. The question can be restated in terms of the location of
the ring. The fact that the shepherd moons move on chaotic trajectories makes this
problem more interesting since, strictly speaking, it breaks the periodicity of
the perturbations. With regards to this, it is worth quoting~\cite{Tiscareno2013},
who points out that, despite the time-varying clumpy and kinky structure, the F
ring core ``maintains over decadal timescales the shape of a freely precessing
eccentric inclined ellipse; the orbital solution formulated to account for
Voyager and other pre-Cassini data~\citep{BoshEtAl2002} has, somewhat
surprisingly, remained a good predictor of the core's position through the
Cassini mission''.

In this paper we address the question of the confinement and location of
Saturn's F ring within an independent (non-interacting) test particle model. We
consider a simplified planar restricted five-body model and follow the dynamics
of a large number of test particles ($\sim 2.5$ millions) up to $6 \times 10^6$ orbital
revolutions of Prometheus, slightly more than 10000 years. The model includes
the gravitational field of Saturn with its second gravitational harmonic $J_2$,
Prometheus, Pandora and Titan; preliminary results appear
in~\citep{BenetJorba2013}.
We find a broad set of test-particle initial conditions that remain trapped
within the region between Prometheus and Pandora. Within this set, there are
initial conditions that remain well-localized and correspond to the more stable
ones with respect to their radial excursions. A projection onto the $X-Y$
space of a snapshot of this stable subset yields a narrow and eccentric ring
with collective alignment,
whose location properties and width can be compared with the observations.

The model we consider is simple in the sense that it does not include
contributions which are important for a detailed realistic description, such
as the $J_4$ gravitational harmonic or the gravitational interaction of other
Saturn moons. The assumption of non-interacting test particles is
equivalent to neglecting any ring particle collisions and self-gravity. While
the assumptions are rather strong, in particular for the
time-span of our integrations, our view is that the physics of the
existence of the F ring, or other phenomena, is not related to matching
very specific parameters, but because certain necessary conditions are
fulfilled. In the present case, the important property is the existence of
phase-space regions where the radial diffusion is strongly suppressed.

The paper is organized as follows: In section~\ref{Sec:Model} we describe of our
general approach and introduce the simple model that we study.
Section~\ref{Sec:Numerics} describes the numerical results obtained, where
we focus first on the test particles that remain trapped up to a maximum
time, and then classify them dynamically according to their stability. The
stability analysis is based on a dimensionless dynamical indicator related
to the radial excursions. {\sl Filtering} the orbits that do
not satisfy a stability condition yields a narrow, eccentric, sharp-edged
ring, whose semi-major axis and width are then compared with the
observations. We relate the accumulation in the semi-major axis of the
ring particles, and the gaps between them, to orbital resonances, which involve
mainly Prometheus outer Lindblad and co-rotation resonances, but not
exclusively. Finally, in section~\ref{Sec:Concl} we summarize
our approach and results.

\section{A simplified model for Saturn's F ring: the scattering approach}
\label{Sec:Model}

A complete description of the dynamics of Saturn's F ring must contain the
gravitational interactions of Saturn including its flattening, the influence
of all Saturn's moons including in particular the shepherd moons, Prometheus and
Pandora, and the interactions among the ring particles themselves. The latter
involves non-trivial processes associated with physical collisions among the
particles of the ring, such as accretion and fragmentation
processes~\citep{AttreeEtAl2012,AttreeEtAl2014}, which
are particularly relevant because the F ring is located near
the edge of the Roche zone~\citep{CanupEsposito1995}. Needless to say, the
understanding of such a system is difficult. Here, we shall study a simpler
model with the hope that it may clarify some aspects of the confinement of
Saturn's F ring.

Our approach is based on the observation that some test particles eventually
``escape'' from the proximity of the ring, while others remain trapped; the
latter are those we observe. This is the essence of the scattering approach to
narrow rings~\citep{BenetSeligman2000,MerloBenet2007}. Consider the planar
motion for the
ring particles, which are assumed independent massless test particles, i.e.,
mutual collisions are neglected. For a short-range potential moving on a
circular orbit around the central planet, one can prove that there exist stable
periodic orbits in phase space~\citep{BenetSeligman2000}. The linear stability of
such orbits and KAM theory~\citep{delaLlave2001} in a two degrees of freedom
system suffice to show that initial conditions which are close
enough to the stable periodic orbits are dynamically trapped. Initial conditions
beyond certain distance, essentially given by the invariant manifolds of a
central unstable periodic orbit in the vicinity, will escape along scattering
trajectories. The whole region of trapped motion in phase space, which includes
different values of the conserved quantity (Jacobi constant), is of central
interest. In particular, the projection of this region onto the $X-Y$ space
forms a narrow ring; its narrowness follows from the actual (small) region
of trapped orbits in phase space. The ring may be eccentric, since it mimics
the projection of the central stable periodic orbit in the $X-Y$ space, which
in general is non-circular. Non-circular motion of the short-range potential
implies an extended phase-space, due to the explicit
time-dependence of the Hamiltonian, which may produce a richer structure,
e.g., multiple components and arcs~\citep{BenetMerlo2008,BenetMerlo2009}. For
small enough perturbations, the stable organizing centers of the dynamics are
preserved, which in turn preserves the region of trapped motion in
phase space that sustains the ring. The scattering approach is, in that sense,
robust~\citep{BenetSeligman2000}.

The starting point of our model~\citep{BenetJorba2013} is to consider the
motion of an ensemble of
non-interacting massless test particles, defined by the initial conditions
which are in the proximity of the location of the F ring. Their time evolution
shall determine whether a test particle remains (dynamically) trapped and hence
belongs to the ring, or if it simply escapes away. Certainly, this starting
point entails a number of non-realistic assumptions, which lead
to some simplifications. Considering non-interacting test particles allows us
to treat each particle independently, i.e., we may consider a one-particle
Hamiltonian; different initial conditions represent different particles. In
doing this, we neglect any effects due to mutual collisions or related to
their actual shape and size. These properties are important for a detailed
understanding of the fine structure and life-time of the
ring~\citep{Poulet2000,MurrayEtAl2008,CharnozEtAl2009}. We shall also
disregard the influence of the whole ring in the motion of any of the major
bodies or of the particles of the ring; this is tantamount of having massless
test particles. We assume for simplicity that the motion of all bodies takes
place in the equatorial plane of Saturn. These assumptions allow us to
consider a planar restricted $(N+1)$-body problem, where a test particle is
influenced by the motion of $N$ massive bodies.

The model is naturally divided into two parts. First, the motion of the
$N$-interacting massive bodies is given by the many-body Hamiltonian
\begin{equation}
\label{eq:Hn}
{\cal H}_N = \sum_{i=0}^{N-1}\frac{1}{2 m_i}(P_{x_i}^2+P_{y_i}^2)
  - \sum_{i=1}^{N-1} \frac{{\cal G} m_0 m_i}{R_{i,0}}
  \Big(1 + \frac{J_2}{2} \frac{R_S^2}{R_{i,0}^2} \Big)
  - \sum_{1\le i<j}^{N-1} \frac{{\cal G} m_i m_j}{R_{i,j}}.
\end{equation}
Here, $\vec{R_i}$ is the position vector of the $i$-th body with respect to
the origin of an inertial frame, $\vec{P_i}$ denotes the canonically
conjugated momentum and $m_i$ its mass. We denote by
$R_{i,j}=| \vec{R}_i-\vec{R}_j |$ the relative distance between two bodies
and use the convention that $i=0$ represents Saturn, and its moons are ordered
increasingly with respect to their nominal semi-major axis.

In our calculations we use Saturn's mass $m_0=5.68319\times10^{26}\,{\rm kg}$
as the unit of mass~\citep{Jacobson2006},
Saturn's equatorial radius $R_S=60268.0\,{\rm km}$ as the unit of
distance~\citep{Seidelmann2007}\footnote{\label{footnote1}Note that
$R_S$ corresponds to the 1 bar surface equatorial radius
of Saturn~\citep{Seidelmann2007}, and was naively retrieved from
the JPL-Horizons Physical Data and Dynamical Constants table for the
planets~(\url{http://ssd.jpl.nasa.gov/?planet_phys_par}). This value was
used to scale out the semi-major axis (obtained from the $J_2$ first order
corrections and the observed mean motion) to set-up our numerical
calculations. This value is not the standard one used in
the planetary rings literature; the latter is $R_S'=60330\,{\rm km}$
and is also the nominal value used to compute the zonal harmonics of
Saturn~\citep{NicholsonPorco1988,Jacobson2006}, in particular, the
value for $J_2$ that we use. Below, whenever we convert a distance
back to km for comparisons with the actual data, we use for the
conversion factor $R_S'=60330\,{\rm km}$ used in the literature.
The reader is warned about this subtle inconsistency.}, and Prometheus period
$\, T_{\rm Prom} = 0.612986\,{\rm days}$~\citep{Seidelmann2007} as
$2\pi$ units of time (i.e., Prometheus' mean motion is unity). In these units,
the numeric value of the gravitational constant is ${\cal G}\cong 12.311...$.
In Eq.~(\ref{eq:Hn}) we have included the first zonal gravitational coefficient
of Saturn, $J_2=16290.71\times 10^{-6}$~\citep{Jacobson2006}, since we are
interested in somewhat long-time integrations. Higher-order zonal harmonics
have been ignored, since they do not create new resonances and slow down
significantly the numerical calculations; yet, these corrections are important
for realistic models~\citep{RennerSicardi2006,2014ElMoutamidEtAl}. As a
side remark, we
mention that including the flattening of Saturn actually allows to apply KAM
theory in the context of the scattering approach, since the shift in the
frequencies lifts up the degeneracy of the dominating two-body Kepler problem.

The second part of the model is related to the motion of the
test particles. The Hamiltonian for the massless test-particles reads
\begin{equation}
\label{eq:Hrp}
{\cal H}_{\rm{rp}}(t) = \frac{1}{2}(p_x^2+p_y^2) - \frac{{\cal G} m_0}{r_0(t)}
  \Big(1 + \frac{J_2}{2} \frac{R_S^2}{r_0^2(t)} \Big)
  - \sum_{i=1}^{N-1} \frac{{\cal G} m_i}{r_i(t)} ,
\end{equation}
where $r_i(t)=|\vec{r}-\vec{R}_i(t)|$ is the relative distance of the test
particle and the
$i$-th body, $\vec{r}$ denotes the position vector of the ring particle and $\vec{p}$
its momentum. Because of the explicit appearance of time in Eq.~(\ref{eq:Hrp}),
the energy of a test particle is not conserved.

What massive bodies shall we include in a simplified model for the F ring?
Clearly, we must include the shepherd moons Prometheus and Pandora, which are known
to influence the dynamics of the ring, though they do not confine it completely
\citep[see][]{Esposito2006}. Since the masses of the shepherd moons are exceedingly
small, their influence on the test particles is essentially local: Test particles
in the region between the shepherds, experiencing the gravitational field of
Saturn and the shepherds, move along precessing Kepler ellipses which are
locally perturbed by the shepherds. The time scale for the test particles to
escape is extremely large, if finite at all. We shall include the interaction
of another Saturn's moon with the hope that it helps promoting radial
excursions and thus escapes; the model is then a planar restricted
$(4+1)$-body problem.

There is no obvious choice for the third moon in the model; see
Table~\ref{tableParams} for a comparison of the orbital parameters of the
main candidates and the F ring. Two obvious candidates are Titan and Mimas:
Titan is the most massive moon of Saturn's satellite system,
$M_{\rm Titan} \approx 2.3669\times 10^{-4} M_S$, and moves on a slightly
eccentric orbit, but it is located rather far away from the ring,
$a_{\rm Titan} = 1221.47\times 10^3$~km.
In turn, Mimas is the major moon of Saturn closest to the F ring. Moreover,
Mimas could play an important role, since Pandora is close to a $3:2$
co-rotation eccentric resonance with Mimas. Its
orbital eccentricity is smaller but comparable to that of Titan. Yet, its mass
is rather small, $M_{\rm Mimas}\approx 6.6\times10^{-8}\, M_S$. The force
exerted by Titan on a test particle at the nominal semi-major axis of
the F ring, $a_{\rm F ring}=140221.6\,{\rm km}\approx 2.324243... \, R_S'$,
\citep{BoshEtAl2002}, is about $4$ times larger than the force
exerted by Mimas, when Titan is at the location furthest from the particle
and Mimas is at the closest one. In addition,
\citet[][see Appendix B]{CuzziEtAl2014} remark that numerical integrations
without Titan shift the location and order of the Prometheus' outer Lindblad
resonances, which play an important role, as we shall also show. To keep
the model simpler as possible, we
shall consider only the influence of Titan on a precessing Kepler elliptic
orbit. We are thus assuming that Titan's mass and eccentric motion
helps to create instabilities in the region of the ring that promote
escapes from that region, though allowing for the existence of regions
of stable motion. Numerical calculations with and without Titan
show that Titan indeed promotes radial excursions, though the actual
effect is weak; see~\ref{Sec:appendix}.
Due to the small mass ratio between Titan and Saturn and
the shepherds and Titan, we shall simplify the numerics by computing the
precessing motion of Titan due to the gravitational attraction of Saturn
and its $J_2$ coefficient only, and for the shepherds we include the
additional perturbations by Titan. We also consider that Saturn remains at
the origin.

\begin{table}[t]
  \caption{Saturn's moons and F ring data, taken from the JPL Horizons's
  ephemerids web interface.  The semi-major axis and eccentricity of the F
  ring, which we shall consider as the nominal values, are taken
  from \citet[][fit 2 of Table III]{BoshEtAl2002}.
  }
  \label{tableParams}
  \begin{minipage}{\textwidth}
    \begin{tabular}{p{2.5cm} c p{1.0cm}l p{1.0cm}c}
        \\
        \hline\noalign{\smallskip}%
        Object    & $a$ ($10^{3}\,$km) & & $\quad e$  & & m ($10^{19}\,$kg)\\
        \noalign{\smallskip}\hline\noalign{\smallskip}%
        Prometheus  & $139.35$   & & $0.0024$  & & $0.014 $\\
        F Ring
              & $\quad 140.2216$  & & $0.0025$  & &  ? \\
        Pandora     & $141.70$   & & $0.0042$  & & $0.013 $\\
        Mimas       & $185.54$   & & $0.0196$  & & $3.75   $\\
        Encedalus   & $238.04$   & & $0.0047$  & & $10.805 $\\
        Tethys      & $294.67$   & & $0.0001$  & & $61.76  $\\
        Dione       & $377.42$   & & $0.0022$  & & $109.57 $\\
        Rhea        & $527.07$   & & $0.001 $  & & $230.9  $\\
        Titan       & $1221.87$  & & $0.0288$  & & $13455.3$\\
        Hyperion    & $1500.88$  & & $0.0274$  & & $1.08   $\\
        Iapetus     & $3560.84$  & & $0.0283$  & & $180.59 $\\
        Phoebe      & $12947.78$ & & $0.1635$  & & $0.8289 $\\
        \noalign{\smallskip}\hline\noalign{\smallskip}%
    \end{tabular}
  \end{minipage}
\end{table}

We are interested in the existence of regions of trapped motion for test
particles between the orbits of the shepherd moons. The
conditions for escape are defined as follows: A test particle escapes if it
leaves the region defined by the orbits of the shepherds, i.e., if it
is not located within the region defined by the innermost
radial position of Prometheus and the outermost of Pandora; these distances
are defined from their nominal locations; c.f. Table~\ref{tableParams}.
In addition, we consider that a test particle collides with a shepherd moon
if it is located within Hill's radius of the moon, that is,
$r_i < R_{H_i} = a_i (m_i/ (3 m_0))^{1/3}$ (with $i=1,2$ for Prometheus
and Pandora respectively). In this case, the test particle shall also be
treated as an {\it escaping} particle, since such an event corresponds
to a physical collision with one of the shepherds~\citep{Ohtsuki1993}. In
either case, the integration of the orbit is terminated and the test particle
is disregarded. On the other hand, if the test particle
does not fulfill any of these requirements before the end of the numerical
integration at $\sub{t}{\rm end}$, the test particle is said to be a trapped
particle and, in that sense, a particle of the ring. We shall see below that
an additional dynamical criterion related to the stability properties of the
orbit must be imposed, which accounts for the possibility of a future
escape. We rely on numerical simulations to study the dynamics of the
test particles.


\section{Numerical results}
\label{Sec:Numerics}

The results presented below were obtained using a high-order Taylor's
integration method (maximum order of the Taylor expansion was $28$) for the
numerical integration~\citep{Jorba2005}. The absolute error of
the largest of the two last terms of all Taylor expansions in every
integration step, which is used
to determine each step size of the integration, was fixed to
$\epsilon=10^{-21}$. The accuracy of the integrations is such
that the energy and the angular momentum of Titan's Kepler motion,
in absolute terms, is conserved throughout the whole integration below
$2\times 10^{-14}$.

\subsection{Initial conditions and trapped orbits}

We first define the initial conditions for this problem. Since
the model assumes independent test particles, each particle is represented by
an independent initial condition of the Hamiltonian~(\ref{eq:Hrp}). The initial
conditions for each test particle are fully characterized by its semi-major
axis $a$, its eccentricity $e$, and two angles that define the orientation of the
initial ellipse $\omega$ with respect to an arbitrary direction, the argument of
pericentre, and the actual position $M$ along the ellipse, the mean anomaly.
We shall focus in the phase-space region defined by $a\in[2.318,2.345]$,
$e\in[0,1.45\times 10^{-3}]$, and $\omega, M \in [0,2\pi]$. This region, when
projected onto the coordinate space, spans the (planar) region between the
orbits of Prometheus and Pandora. Hence, the orbital elements of the particles
of Saturn's F ring are included in this phase-space region (under the
assumption of planar orbits), though we emphasize that this region is not
restricted only to such orbital elements. As
we shall show, the dynamics of this system {\it selects}, through the escaping
mechanism and a further stability condition, a specific subset of these
initial conditions.

\begin{figure}
  \centering
  \includegraphics[width=0.7\linewidth]{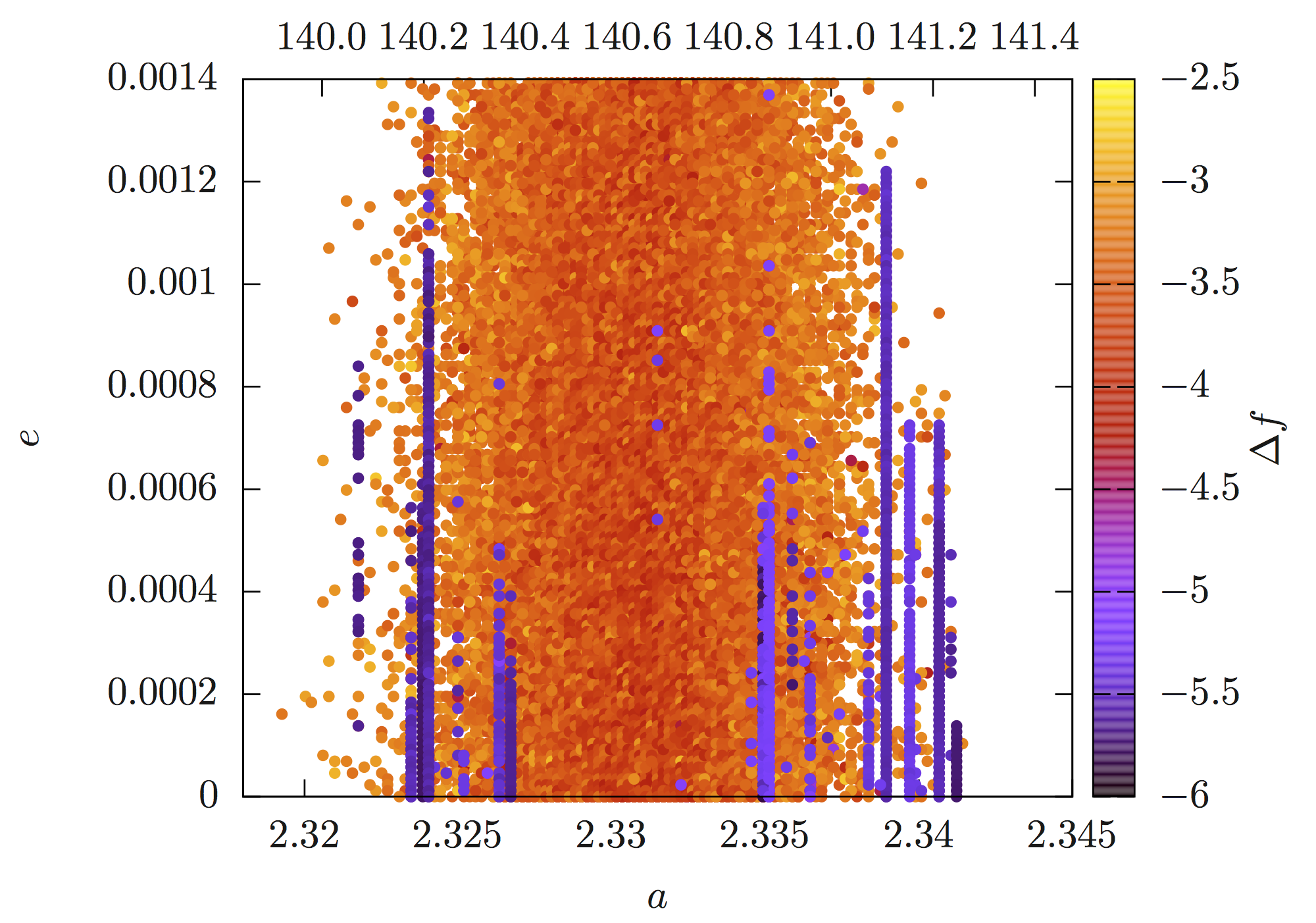}
  \caption{\label{fig1}
      Projection onto the semi-major axis $a$ and eccentricity $e$ plane of the
      initial conditions of test particles that remain trapped for at least
      $\sub{t}{\rm end}=2.4\times 10^6 \, T_{\rm Prom}$.
      The color code is the stability index $\Delta f$ associated with the mean
      motion in logarithmic scale (see text). Notice that there are two localized
      regions in the $a-e$ plane, where the particles exhibit an enhancement of
      the stability index (blue-purple stripes). The lower $a$-axis
      is expressed in Saturn radius units $R_S$ and the upper
      one is expressed in $10^3$~km.
      }
\end{figure}

With regards to the initial conditions of the massive bodies, these will be
fixed to an arbitrary but fixed value for all integrations. The case we have
considered is rather artificial, all moons are initially aligned at their
pericentre, with their orbital elements corresponding to the nominal ones
(see footnote~\ref{footnote1}). As
we shall show below, the results for this particular situation have a good
correspondence with the observations; this illustrates the robustness of the
proposed approach.

In Figure~\ref{fig1} we display the initial conditions for the test particles
that remain trapped for at least $\sub{t}{\rm end}=2.4\times 10^6$ periods of Prometheus,
projected onto the semi-major axis $a$ and eccentricity $e$ plane. In these
simulations, the orientation of the initial Kepler ellipses was varied
considering 10 equally spaced values for $\omega$ for $M=0$; the semi-major axis
and the eccentricity was set on a grid of $256\times 256$ points for each
value of $\omega$. Leaving aside the color code for the moment, which will be
explained below, the results shown in this figure indicate that the
trapped particles occupy a rather wide and connected region for both
$a$ and $e$. This statement holds for other initial values of $\omega$ and $M$.

These results seem deceiving at first sight: In general, the inclusion of
Titan in the model promotes more extended radial excursions of the test
particles. These radial excursion may induce collisions with the
shepherds or escape from the region between the shepherds. However, for
the integration times considered so far, the result is a wide region in
the semi-major axis occupied by the trapped test particles. That is,
Titan does not confine the trapped test particles to a narrow region in
$a$, where a (narrow) ring could be located and compared with the
observations.

The fact that Titan promotes more extended radial excursions without
clearing out efficiently (in a short-time scale) the region between the
shepherds motivates to explore the stability properties of the
motion of the test particles~\citep{BenetJorba2013}. We shall do so
considering a variation of the frequency analysis~\citep{Laskar1990}. To be
more specific, we characterize the variation of a given frequency, or an
action, along the whole integrated orbit using a stability index defined by
the standard deviation of the frequency (or action) normalized by its
average value. In
particular, in Figure~\ref{fig1} we considered the stability index
$\Delta f = \sigma_f/\overline{f}$ associated with the mean motion $f$. In
terms of $\Delta f$, periodic or quasi-periodic motion implies $\Delta f=0$.
Therefore, large-enough non-zero values of $\Delta f$
indicate clear departures from stable or quasi-periodic motion.

\begin{figure}
  \centering
  \includegraphics[width=0.7\linewidth]{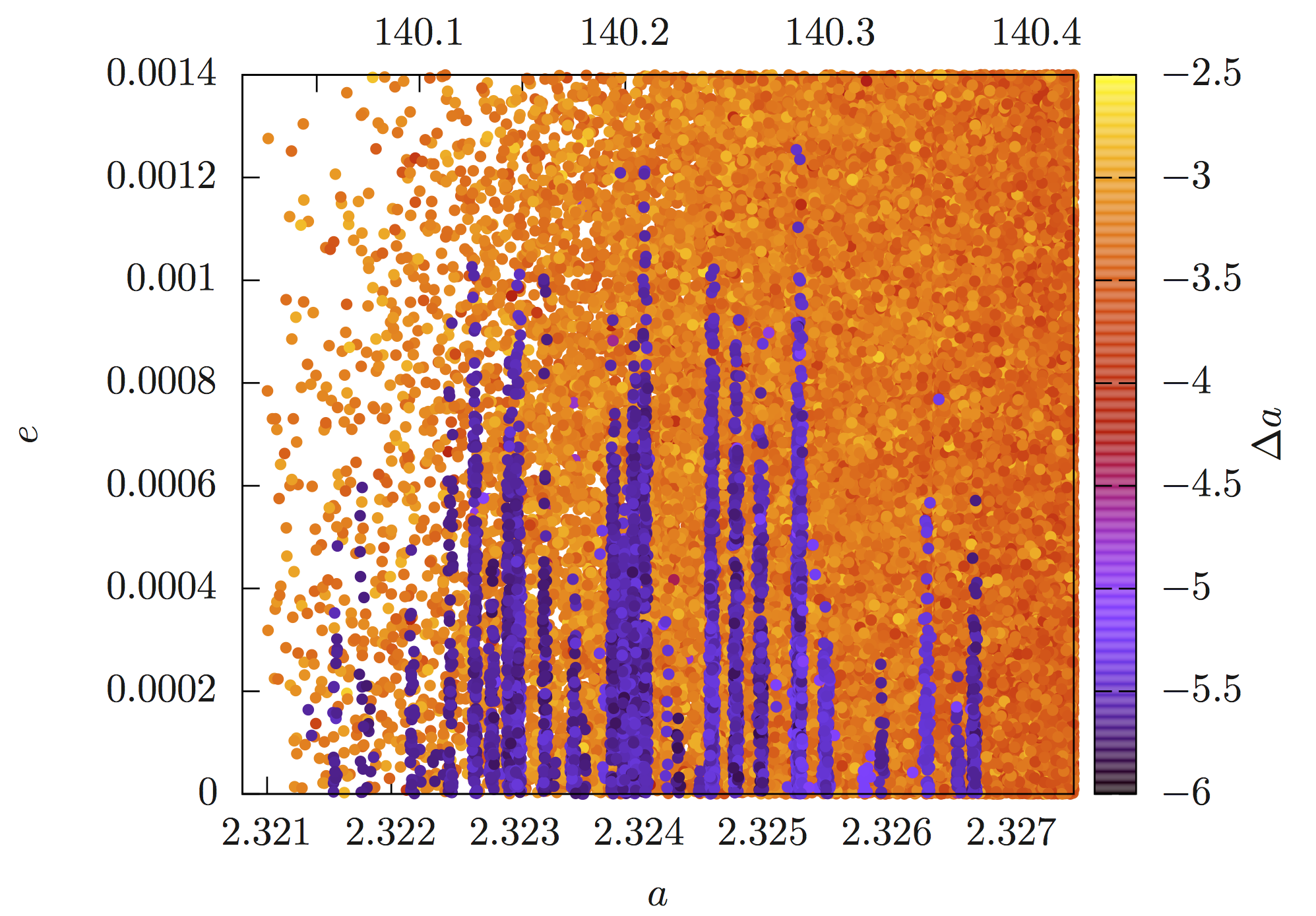}
  \caption{\label{fig2}
  Projection onto the semi-major axis $a$ (close to the region where
  the F ring is observed) and eccentricity $e$ plane of the
  initial conditions of test particles that remain trapped at least up
  to $\sub{t}{\rm end}=3.2\times 10^6 \, T_{\rm Prom}$. For this figure
  the whole $M$ and $\omega$ domains were considered. The stability index
  has been coded with respect to $\Delta a$, the normalized standard
  deviation of $a$, in logarithmic scale. Note that the scales differ
  from those in
  Figure~\ref{fig1}, and include the region with Saturn's F ring is
  actually observed. The lower horizontal scale is expressed
  $R_S$, and the upper one is given in $10^3$~km.}
\end{figure}

From the numerical integration for each test particle, we computed the main
frequencies every 200 revolutions of Prometheus, using a collocation method
for the Fourier analysis~\citep{Gomez2010}. The color code used in
Figure~\ref{fig1} corresponds to $\log_{10}(\Delta f)$. Taking into account
the color code, we note the appearance of localized stripes in the
semi-major axis at certain locations, around $a\sim 2.325$, and
$a\sim 2.340$.
These locations are particularly stable in terms of $\Delta f$. Below we
shall show, using longer numerical integrations, that a large proportion of
particles whose stability index is in the orange-yellow region are escaping
orbits, while a minor quantity behaves alike in the blue--purple region.

\begin{figure}
  \centering
  \includegraphics[width=0.7\linewidth]{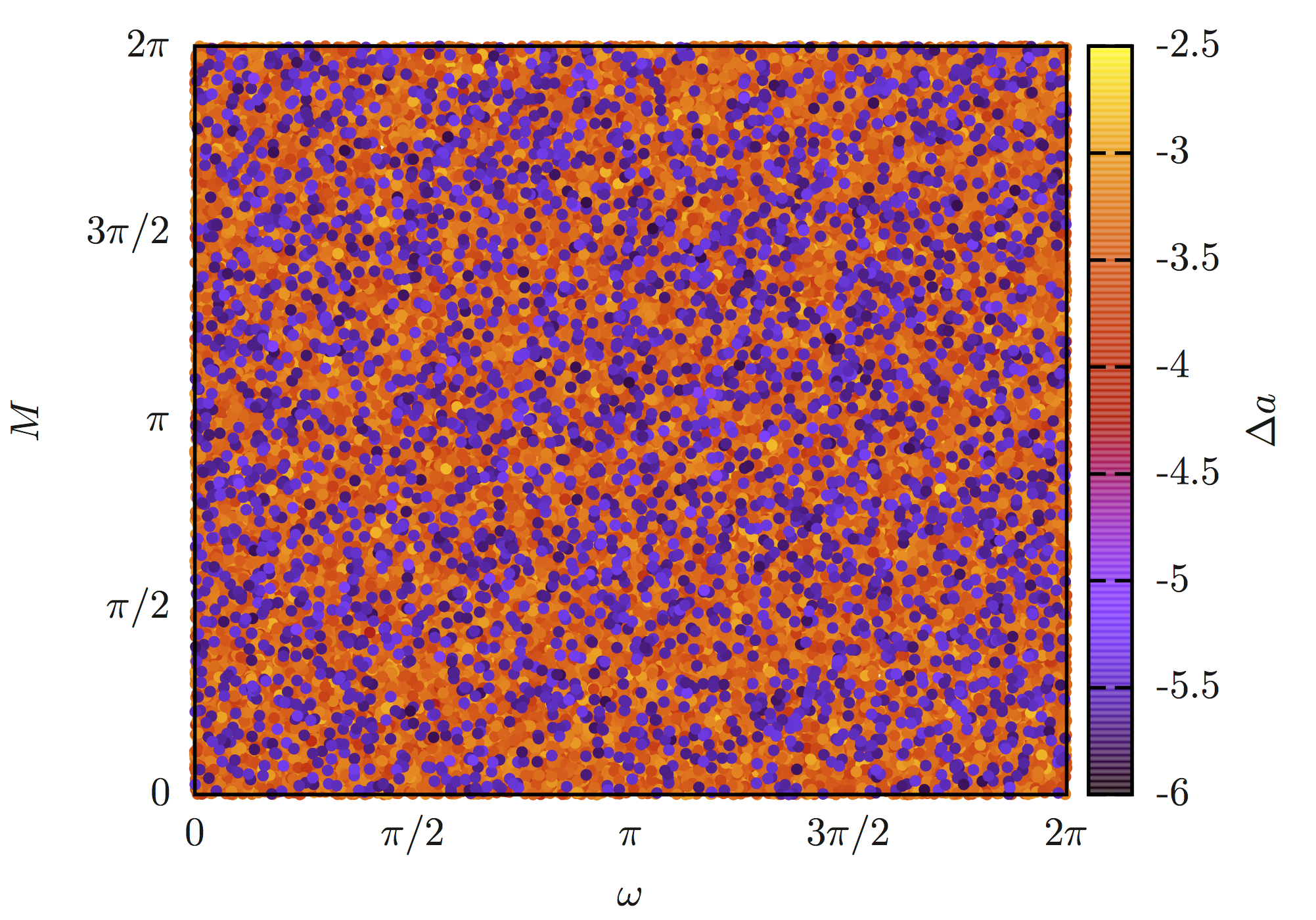}
  \caption{\label{fig3}
  Projection onto the space of angles $\omega-M$ of the same initial conditions
  depicted in Figure~\ref{fig2}. The distribution of stable trajectories
  seems quite homogeneous, indicating that that the stability index
  $\Delta a$ (in logarithmic scale) is independent of the angular
  components of the initial conditions.}
\end{figure}

It is interesting to note that the first group of blue--purple stripes mentioned
above, located around $a\sim 2.325$, is close to the nominal semi-major axis of
Saturn's F ring; cf. Table~\ref{tableParams}. We shall focus on this region
from now on, due to its correspondence with the nominal orbital data of
Saturn's F ring.
We note that in Figure~\ref{fig1}, as well as in Figures~\ref{fig2},
\ref{fig4} and \ref{fig6} displayed below, we have included an upper
scale for the radial distances given in km, which may be used to compare
with the nominal data. For the conversion we use $R_S'=60330$~km,
since this value is used to obtain the nominal value for $a_{\rm F ring}$
during the data reduction and fitting
processes~\citep[see][]{BoshEtAl2002,Cooper2013,CuzziEtAl2014}. Yet,
our main interest is not on the correspondence to this value
but on a generic mechanism that yields a narrow ring.
We note though that, using this factor yields a better comparison
of our results with the nominal data than using the value of
$R_S$ from~\citet{Seidelmann2007}
which we used to scale out the physical units in our numerical
simulations. The differences in the scale are below $0.2\%$ due to
the difference in the $R_S$ values. For
$a\sim 2.324$, this difference amounts to $\sim 140$~km.

Figure~\ref{fig2} illustrates the projection of the
trapped orbits around this region for longer integrations
($\sub{t}{\rm end}=3.2\times 10^6 \, T_{\rm Prom}$); the initial
conditions are taken from the whole $M$ and $\omega$ domains. In
this case, the stability index has been computed with respect to $\Delta a$,
i.e., with respect to the normalized standard deviation of the semi-major axis
of the test-particle's orbit, which was calculated
using an averaging extrapolating method to
achieve rapid convergence~\citep{LuqueVillanueva2014}. Clearly, $\Delta a$
is a measure of the radial excursion of the test particle. Again, groups of
blue--purple stripes appear localized around certain values of the semi-major
axis. Regarding the angular correlations of the trapped test particles,
Figure~\ref{fig3} shows that the initial angles $M$ and $\omega$ do not display any
prominent structure with respect to the stability index.

\begin{figure}
  \centering
  \includegraphics[width=0.7\linewidth]{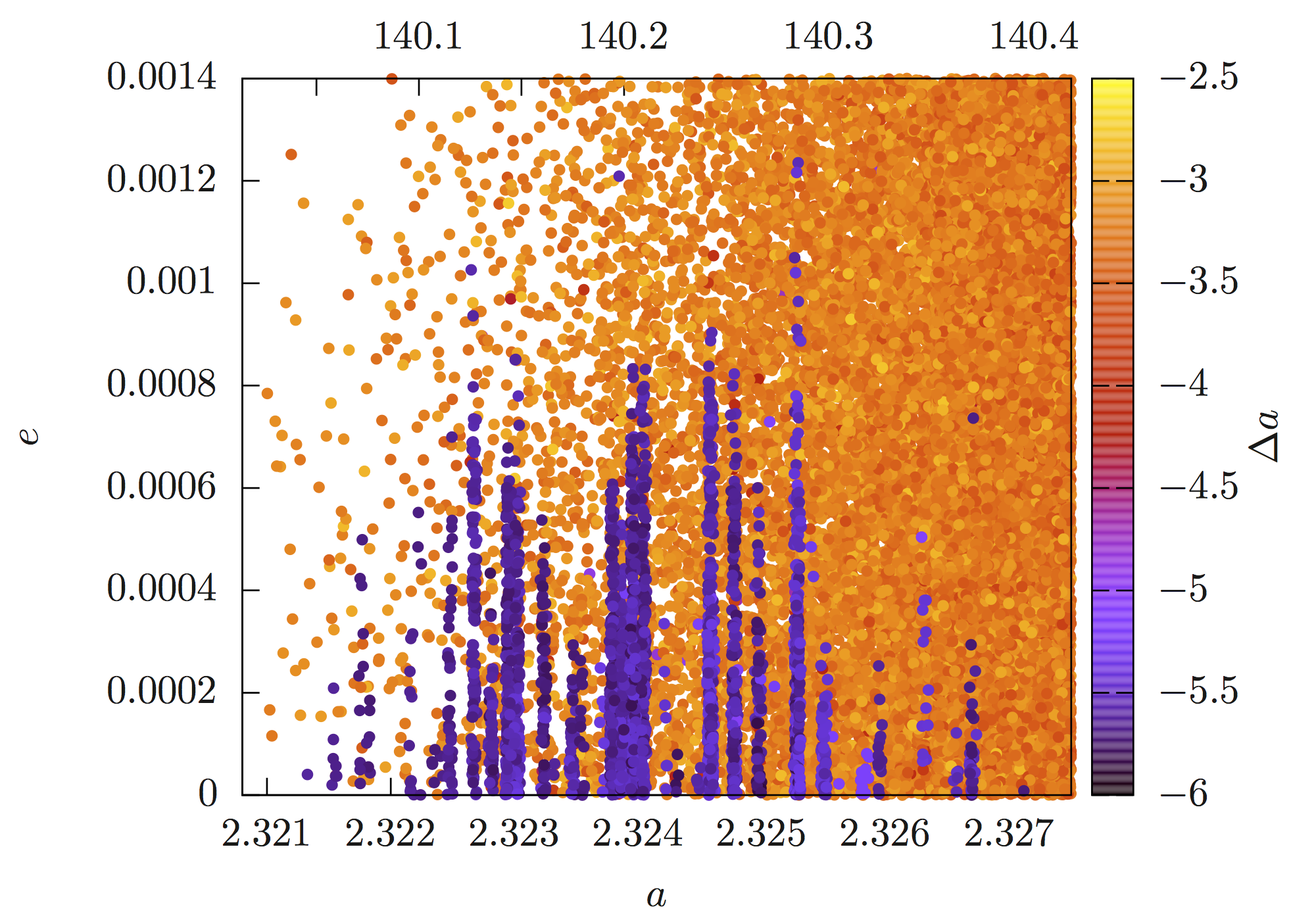}
  \caption{\label{fig4}
  Same projection as in Figure~\ref{fig2} for test particles that stay trapped
  at least up to $\sub{t}{\rm end}=6.0\times 10^6 \, T_{\rm Prom}$. Notice the
  appearance of a white background indicating that some test particles have
  escaped. The majority of the initial conditions of the escaping test particles
  have stability indexes in the orange-yellow region.
  The lower horizontal scale is expressed
  $R_S$, and the upper one is given in $10^3$~km.}
\end{figure}

In Figure~\ref{fig4} we display the projection into the $a-e$ space of the
initial conditions that remain trapped at least up to
$\sub{t}{\rm end}=6.0\times 10^6 \, T_{\rm Prom}$. The structure resembles
that of Figure~\ref{fig2}, except that many initial conditions that were
trapped in Figure~\ref{fig2} have now escaped. The vast majority
of the escaped particles correspond to test particles in the orange-yellow
region of the stability index $\Delta a$ displayed in Figure~\ref{fig4}. An
example of this can be easily noticed by the white
regions appearing on the middle-left of Figure~\ref{fig4}, which were before
occupied mainly by orange-yellow dots.

The last observation confirms the naive expectation that only the most stable
particles will remain trapped, in the sense that their radial excursions
are strongly limited~\citep{BenetJorba2013,CuzziEtAl2014}. While
this statement may sound trivial, we
emphasize that the Hamiltonian~(\ref{eq:Hrp}) is explicitly time-dependent,
and its time dependence reflects the dynamics of the $N$-body Hamiltonian.
Moreover, we recall that the shepherd moons move along seemingly chaotic
orbits~\citep{PouletSicardi2001,FrenchEtAl2003,
GoldreichRappaport2003a,GoldreichRappaport2003b}.

\subsection{Statistical properties of the stability index and rings}

\begin{figure}
  \centering
  \includegraphics[width=0.65\linewidth]{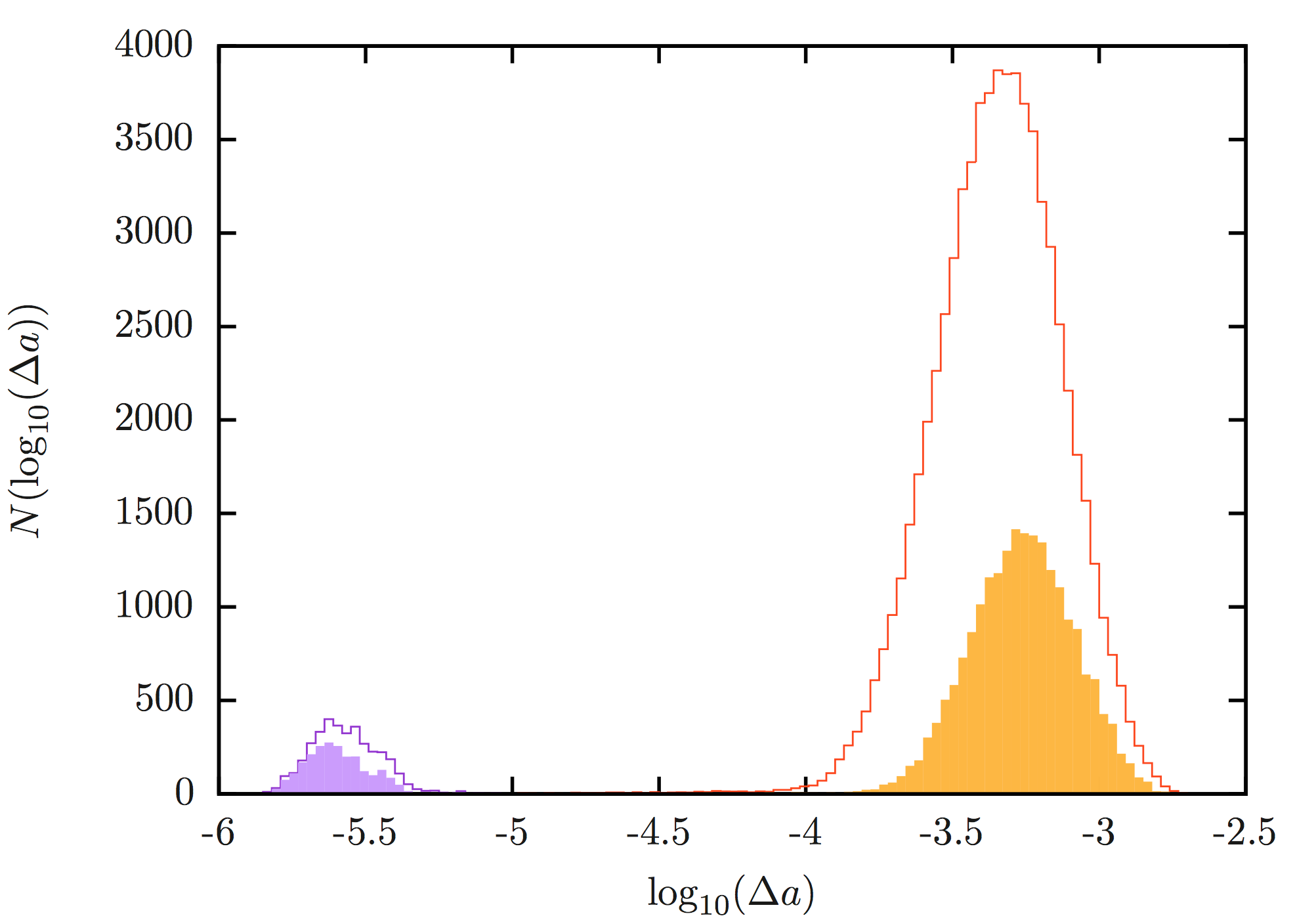}
  \caption{\label{fig5}
  Frequency histogram of $\log_{10}(\Delta a)$ corresponding to the data used in
  Figure~\ref{fig2} for $\sub{t}{\rm end}=3.2\times 10^6 \, T_{\rm Prom}$
  (empty histograms) and to the data of Figure~\ref{fig4} for
  $\sub{t}{\rm end}=6.0\times 10^6 \, T_{\rm Prom}$ (filled
  histograms). Note that the number of initial particles
  with larger stability index (orange-yellow region) shrinks in a larger
  proportion than the region of smaller $\Delta a$ (blue-purple region).
  }
\end{figure}

Figure~\ref{fig5} displays the frequency histograms of the logarithm of
$\Delta a$ for the trapped test-particles considering the two distinct
integration times. The empty histograms corresponds to the data
used in Figure~\ref{fig2} for $\sub{t}{\rm end}=3.2\times 10^6 \, T_{\rm Prom}$,
and the filled histograms to Figure~\ref{fig4} for
$\sub{t}{\rm end}=6.0\times 10^6 \, T_{\rm Prom}$. From Figure~\ref{fig5},
we observe that the trapped test particles clearly display two well-separated
scales according to $\Delta a$. These scales correspond, for a test particle
with $a\simeq 2.324$, to radial excursions in the kilometer scale for
$\Delta a \lesssim 10^{-5}$ (blue-purple region), or from tens to
hundreds of kilometers for $\Delta a \gtrsim 10^{-4}$
(orange-yellow region).

A test particle moves along a
perturbed precessing ellipse with a limited radial excursion. Titan's
eccentric orbit promotes more extended radial excursions
(see~\ref{Sec:appendix}) which may be
amplified by local encounters with the shepherd
moons~\citep[c.f. Figure~8 in][]{CuzziEtAl2014}. These combined perturbations
create conditions which result in a larger change in the
semi-major axis and eccentricity. The net result is a more extended
radial excursion of the test-particle orbit, and hence a comparatively large
value of $\Delta a$, which could lead to the escape of the test particle.
Test particles that somehow experience
these large radial excursion but remain trapped at time
$\sub{t}{\rm end}$, correspond to the orange--yellow points in
Figures~\ref{fig2} and~\ref{fig4}, i.e., to somewhat large values of
$\Delta a$. The blue-violet dots in those figures correspond to test
particles that have
experienced very few or no abrupt changes in their semi-major and
eccentricity, and therefore, the associated $\Delta a$ value remains
very small. Figure~\ref{fig5} shows that the test-particles in the
orange-yellow region are more prone to abandon the region between
the shepherds than those in the purple-blue region.

It is only matter of time before test
particles in the orange--yellow region will eventually escape. This
statement does not preclude that test particles with very small values of
$\Delta a$ (measured at a given time) also escape, though it suggests that
the typical time scale needed for them to escape may be much longer. Note that,
as soon as there is an important radial excursion, the value of $\Delta a$
increases, which reflects the instability of the motion in the radial direction.

\begin{figure}
  \centering
  \includegraphics[width=0.85\linewidth]{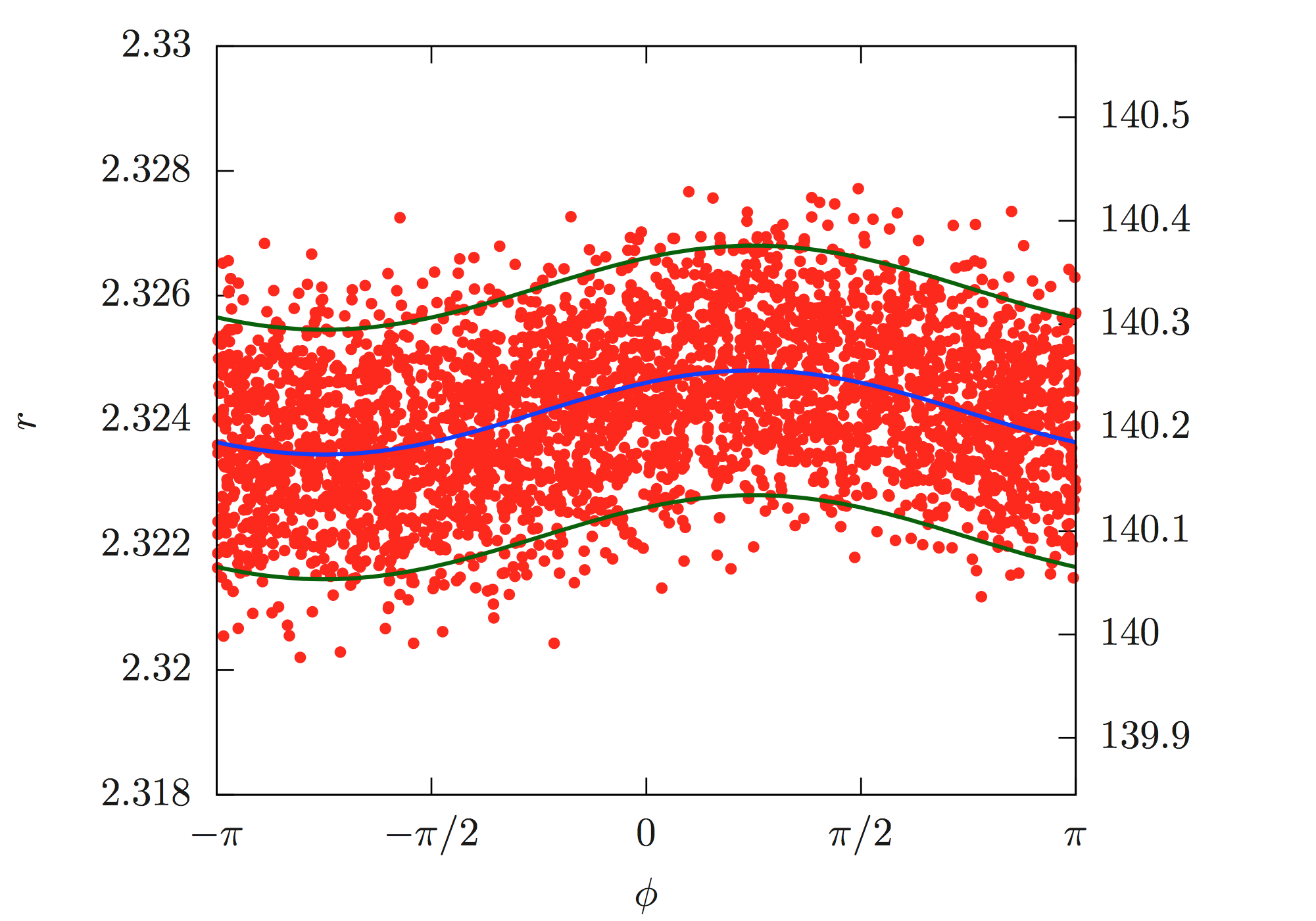}\\
  \includegraphics[width=0.85\linewidth]{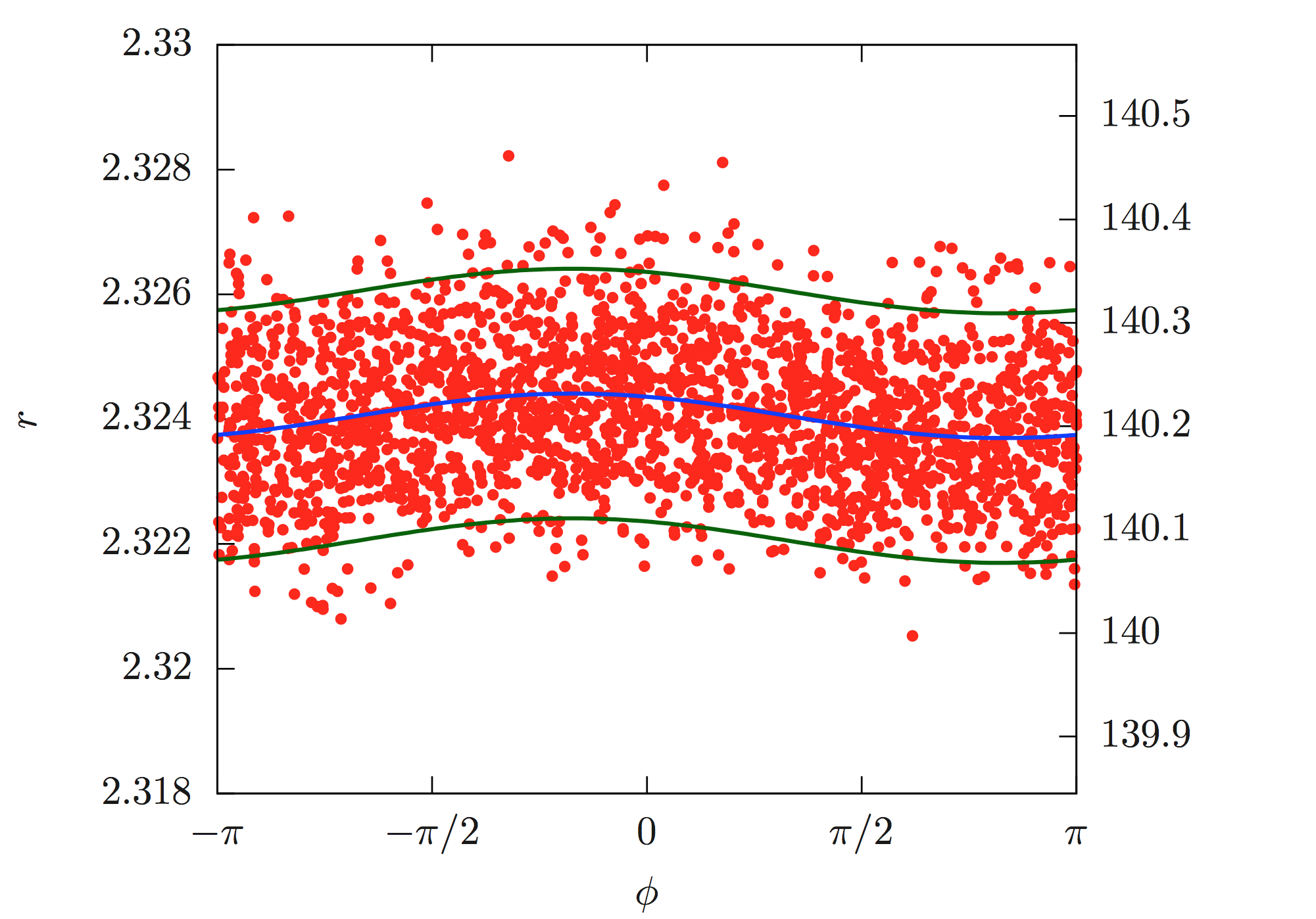}
  \caption{\label{fig6}
  Snapshots of the ring particles, in polar coordinates, after the dynamical
  filtering at (a)~$t=3.2\times10^6\, T_{\rm Prom}$ and
  (b)~$t=6.0\times10^6\, T_{\rm Prom}$. In both
  frames, the blue middle line corresponds to the Keplerian ellipse
  fit considering all the particles of the ring; the outer green lines
  are the same ellipse shifted upwards or downwards by
  $\delta r = 0.002 \approx 120.5\, \textrm{km}$, and permit to obtain a rough
  estimate of the width of the ring. The left vertical
  scale is given in $R_S$ units, and the right one in $10^3$~km.
  }
\end{figure}

The structure of Figure~\ref{fig5} suggests, on dynamical grounds, to filter
the trapped test particles according to their stability index, which
in a way mimics performing longer time integrations. We thus
retain as the actual particles of the ring those whose
stability index is small enough. In particular, we consider that the
ring particles satisfy $\Delta a < 10^{-5}$. Using this dynamical filtering,
in Figure~\ref{fig6} we present the projection of the ring particles onto
the $X-Y$ space using a polar representation; Figure~\ref{fig6}(a) corresponds
to the snapshot taken at $t=3.2\times10^6 \, T_{\rm Prom}$ and~\ref{fig6}(b) to
the snapshot taken at $t=6.0\times10^6 \, T_{\rm Prom}$, respectively. The
ring obtained in each frame is narrow and slightly eccentric.

The rings of Figure~\ref{fig6} have a well-defined collective
orientation, despite of the fact that each ring
particle has independent and randomly chosen initial conditions; this is
another consequence of the dynamical filtering employed, i.e., to
the radial stability criterion. This collective orientation suggests that
the ring displays apse alignment, though we have not proven it; we
shall come back to this later. An estimate of the
instantaneous orbital elements is obtained by fitting a Keplerian ellipse of
the form
\begin{equation}
\label{eq:KepEll}
r = \frac{a(1-e^2)}{1+e\cos(\phi-\phi_0)},
\end{equation}
to all particles of the ring at a given time. For the data in
Figure~\ref{fig6}(a) we obtain
for the semi-major axis $\sub{a}{\rm fit}\approx 2.3241
\approx 140213$~km and the
eccentricity $\sub{e}{\rm fit}\approx 2.9\times 10^{-4}$, using
${\phi_0}_\mathrm{fit}\approx -3\pi/4$. For the data of
Figure~\ref{fig6}(b) we obtain essentially the same semi-major axis
(differences appear beyond the fourth decimal, yielding $140210$~km) and
$\sub{e}{\rm fit}\approx 1.5\times 10^{-4}$ using
${\phi_0}_\mathrm{fit}\approx 2.6$. The
corresponding fits are displayed in Figure~\ref{fig6} by the (middle) blue
lines. The outer green lines correspond to the same fitted ellipse
shifted upwards and downwards by $\delta r = 0.002$, which is approximately
equivalent to $120.5\,\textrm{km}$. These limits include over 90\%
of the ring particles in each snapshot. Hence, these curves provide a rough
estimate of the width of the ring, which is given by
$2\delta r \approx 0.004 \approx 241\,{\rm km}$.

\begin{figure}
  \centering
  \includegraphics[width=0.55\linewidth]{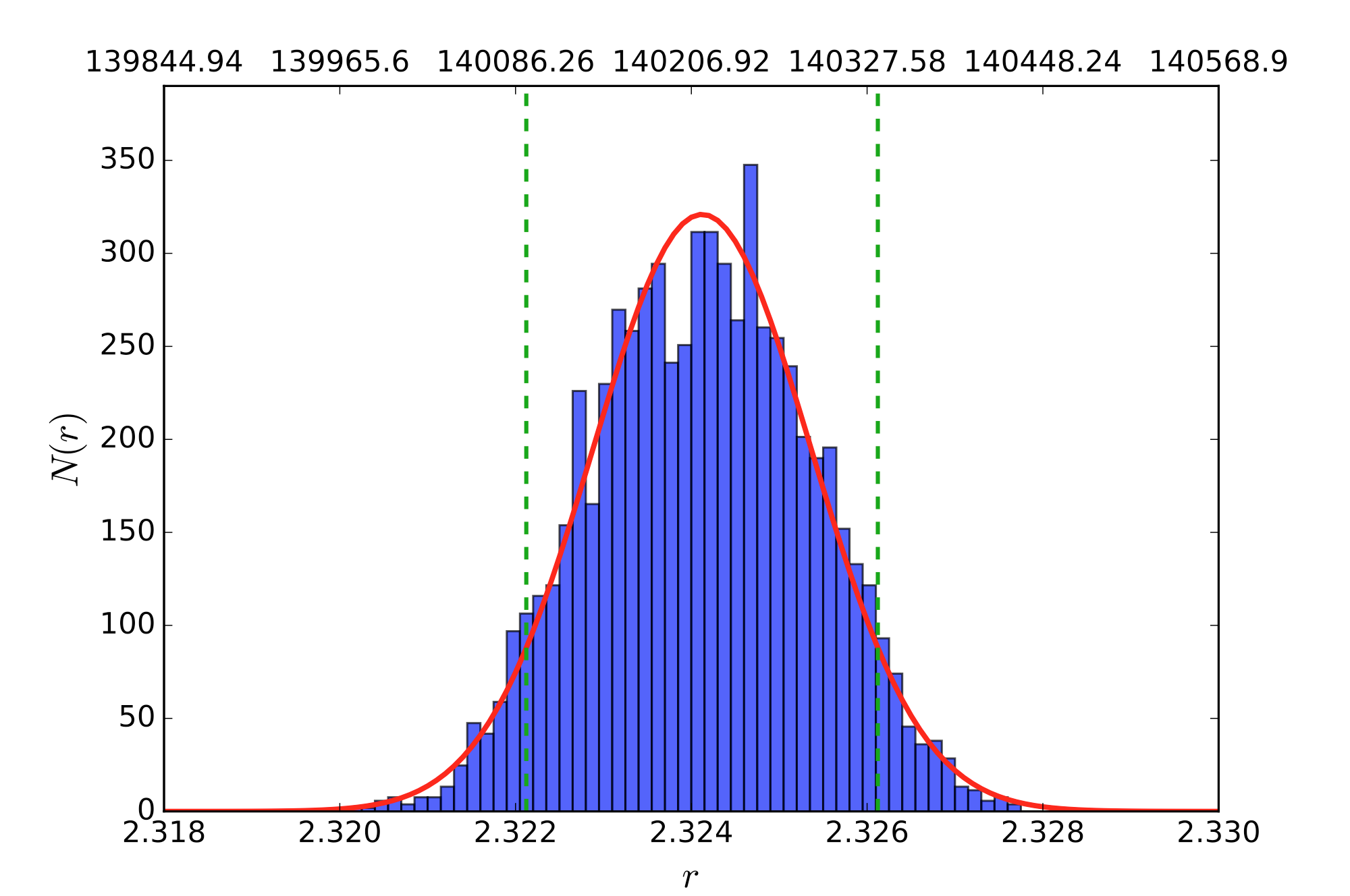}
  \includegraphics[width=0.55\linewidth]{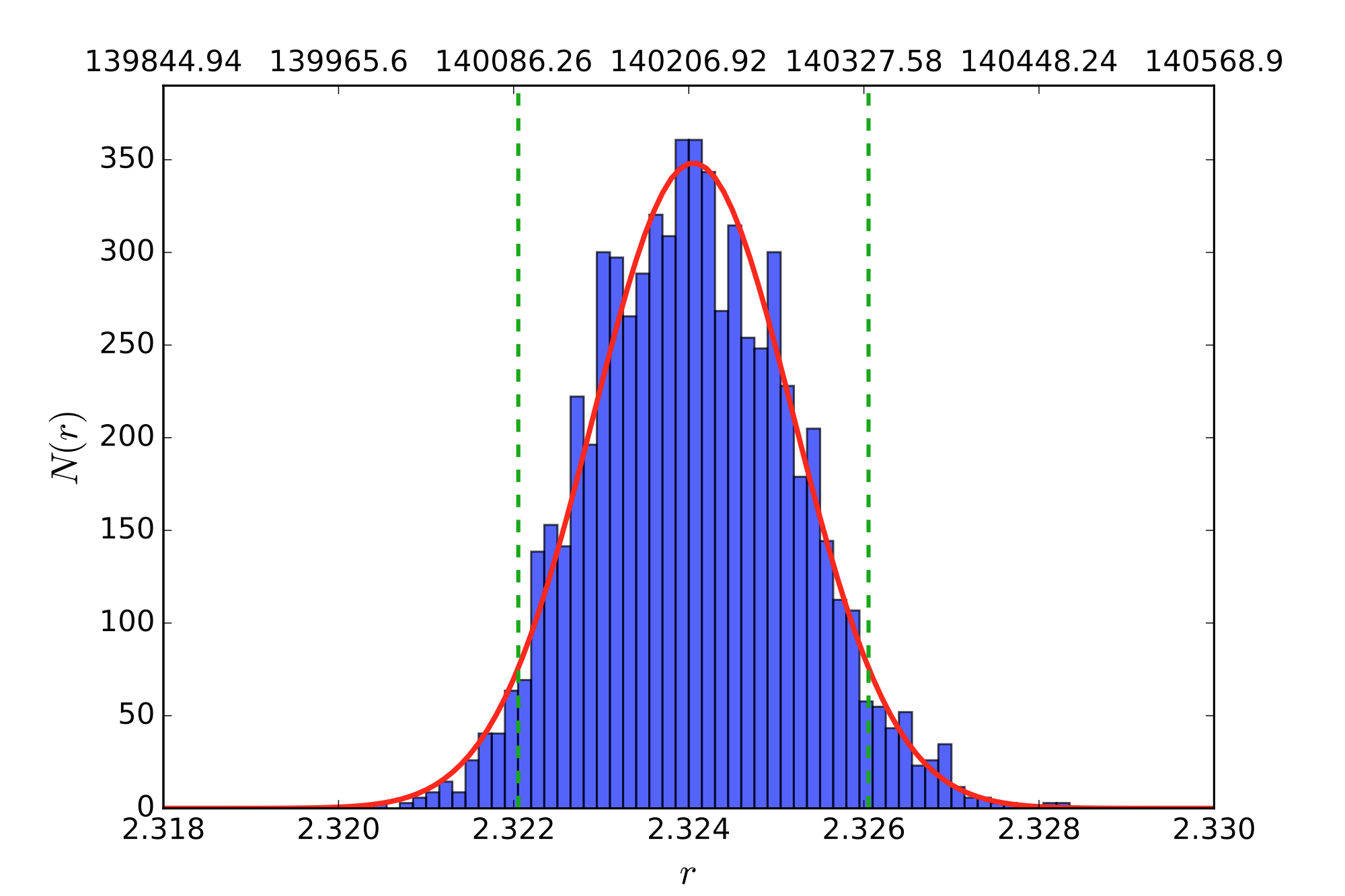}
  \caption{\label{fig7}
  Frequency histograms of the radial positions of the ring particles
  of Figures~\ref{fig6} at (a)~$t=3.2\times10^6\, T_{\rm Prom}$ and
  (b)~$t=6.0\times10^6\, T_{\rm Prom}$. The red curves correspond
  to Gaussian distributions with the same mean and variance of the
  data; these curves are included as a guide for the tails of the
  distribution, which leads us to conclude that the tails have
  at least a Gaussian decay. The vertical lines correspond to the
  shifts by $\delta r$ used to estimate the
  width of the ring (green lines in Figures~\ref{fig6}). The lower
  horizontal axis scale is given in $R_S$, and the upper one in km.
  }
\end{figure}

Note that, the surface density of test particles decreases very fast. In more
quantitative terms, in Figures~\ref{fig7} we present the histogram of
ring particle radial positions corresponding to the rings of
Figures~\ref{fig6}. The red curve in the histograms corresponds to a
Gaussian distribution that uses the mean and the variance computed
from the radial positions. This shows that the surface density of ring
particles decays at least as a Gaussian. It is in this sense that we state that
the ring obtained displays sharp edges.

The semi-major axis of the ring $a_{\rm fit}$ given above is in
correspondence with the observations for Saturn's F ring, namely,
$a_{\rm F ring} = 140221.6\,{\rm km}
\approx 2.324243... R_S'$ \citep{BoshEtAl2002,Cooper2013}.
Differences exist though, that are attributed to the simplicity of the
model. With respect to the width, we note that our definition is similar
to the definition of the core employed by \citet{FrenchEtAl2012}.
Our estimate for the width is consistent with the Voyager 1 results,
though it is about half of the mean-width obtained from the photometric
analysis of Cassini's data, which is wider than Cassini's
occultation data~\citep[see][]{FrenchEtAl2012}. It is also consistent with the
simulations performed by \citet[][cf. figure 18]{CuzziEtAl2014},
and the total extent of the F ring reported in \citet[][see the
discussion of Sect.~6 and Fig.~7]{BoshEtAl2002}.
Yet, the value largely overestimates
the width associated with the core, which is in the range of
$1-40$~km. Note that, while we have obtained an eccentric
ring, the values of the eccentricity obtained are about an order of
magnitude smaller than the observations~\citep{BoshEtAl2002,
AlbersEtAl2012,Cooper2013}. This may be a consequence of neglecting
the larger bodies in the ring, collisions among the ring particles,
or to the overall simplicity of our model.

\begin{figure}
  \centering
  \includegraphics[width=0.56\linewidth]{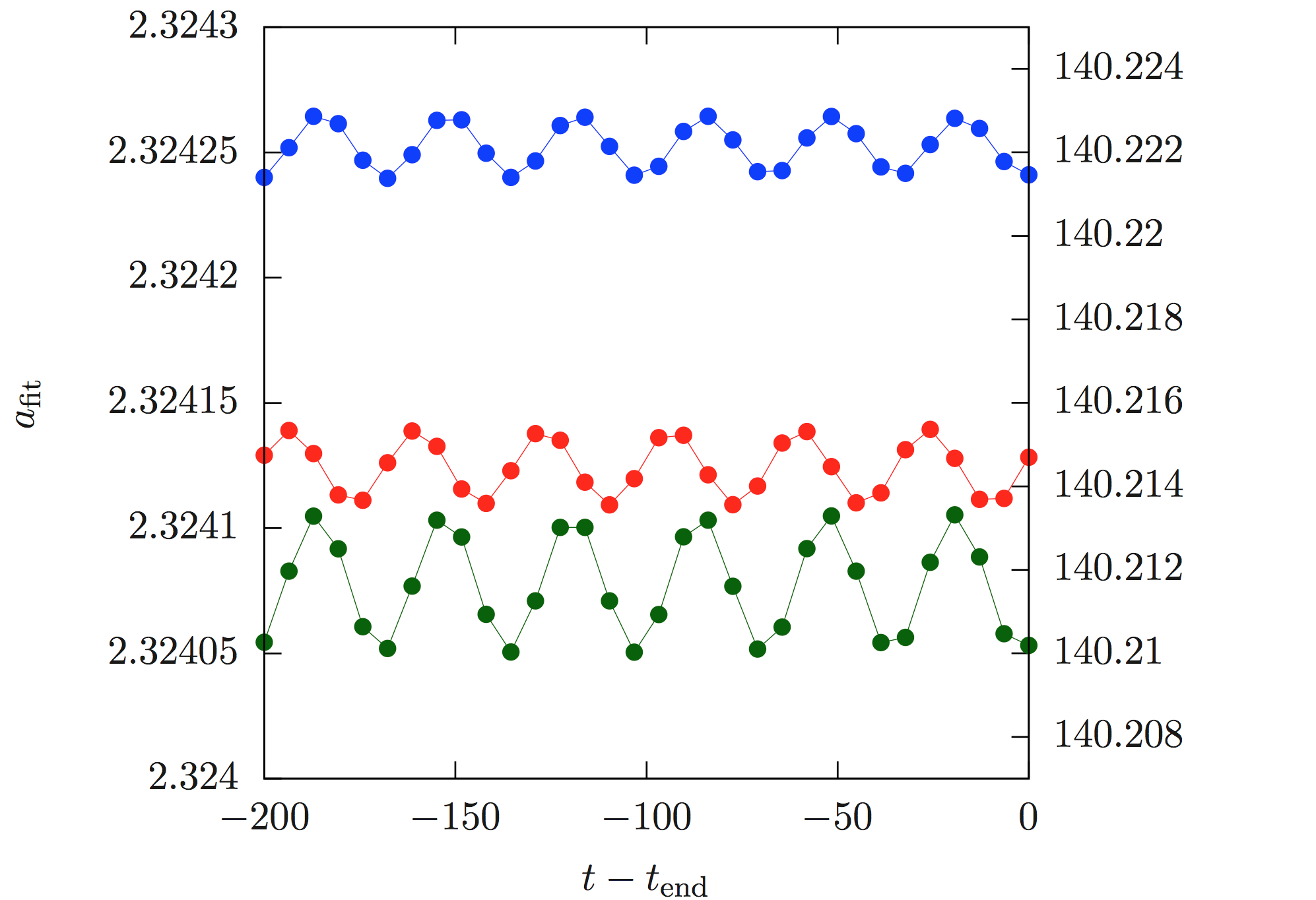}
  \includegraphics[width=0.56\linewidth]{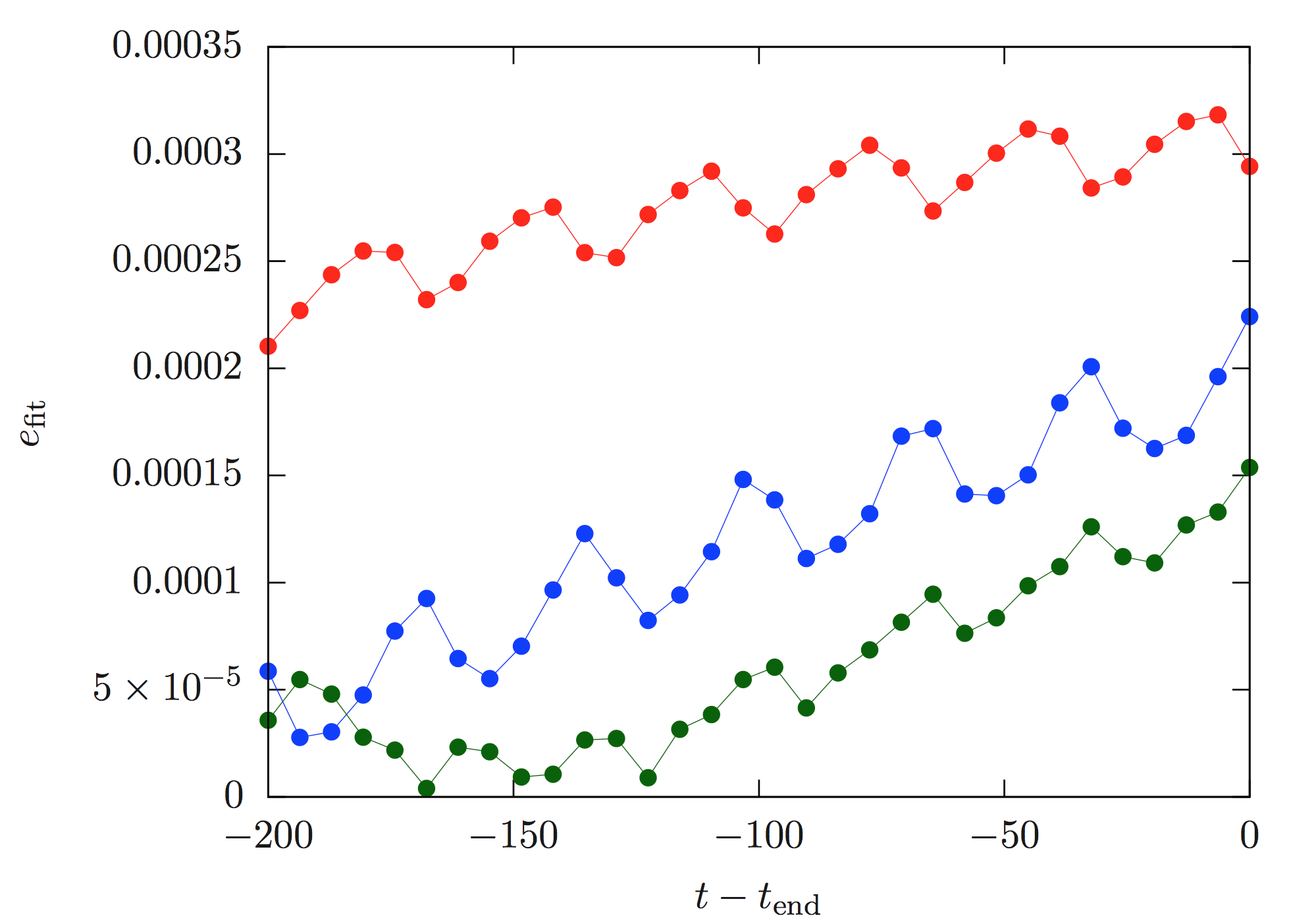}\\
  \includegraphics[width=0.56\linewidth]{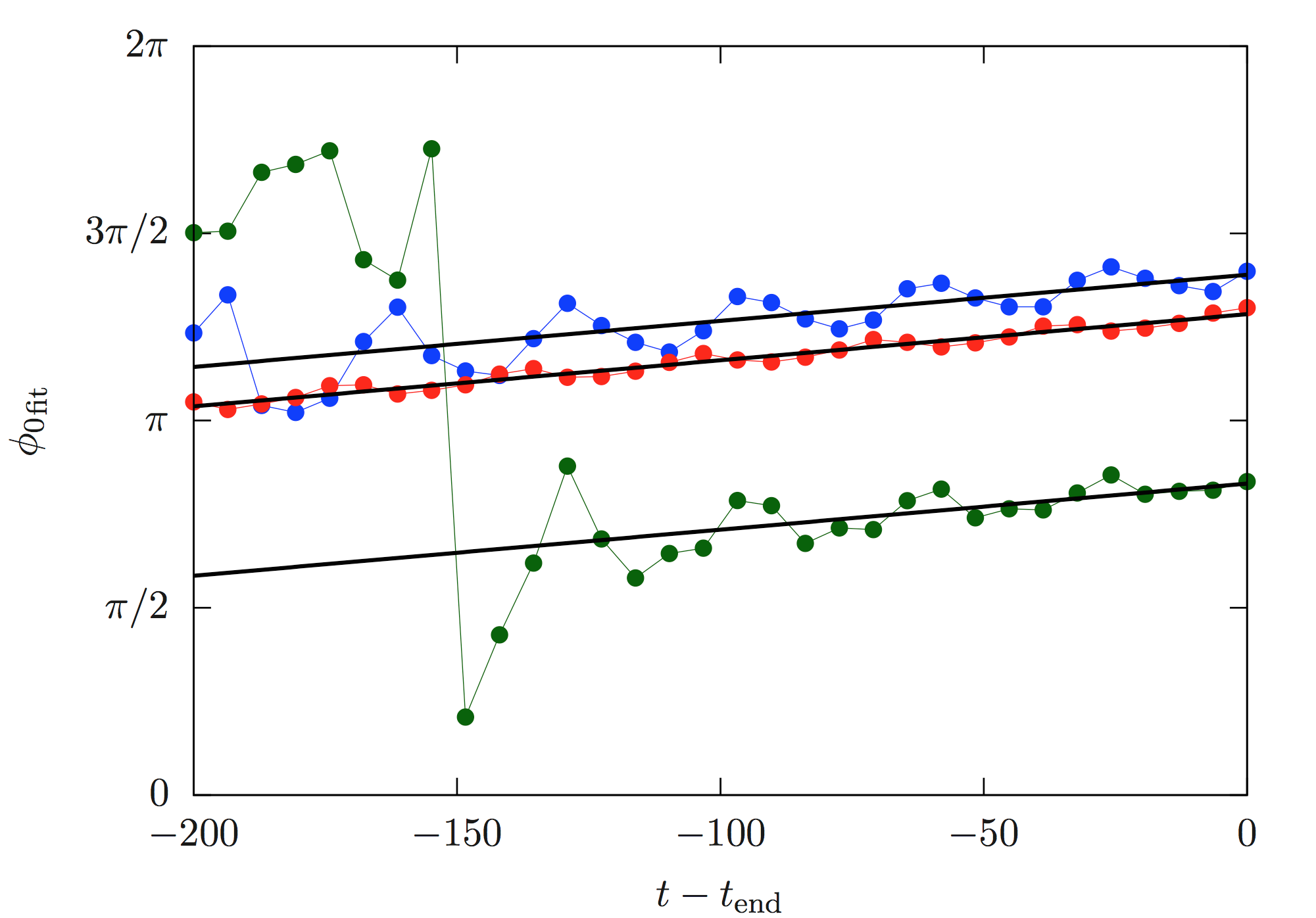}
  \caption{\label{fig8}
  Ring fitting parameters (a)~$a_\mathrm{fit}$, (b)~$e_\mathrm{fit}$ and
  (c)~${\phi_0}_\mathrm{fit}$ as a function of time, spanning
  $200$~time-units ($200/(2\pi)\, T_{\rm Prom}$) close to three
  epochs: $t=1.6\times10^6\, T_{\rm Prom}$ (blue data),
  $t=3.2\times10^6\, T_{\rm Prom}$ (red data) and
  $t=6.0\times10^6\, T_{\rm Prom}$ (green data).
  The left vertical scale of (a) is given in $R_S$
  and the right one in $10^3$~km. The black straight-lines in~(c)
  correspond to a precession rate
  $\dot{\varpi}_\mathrm{fit}\approx 0.003867$~rad/time-units.
  }
\end{figure}

Similar to the fits based on Eq.~(\ref{eq:KepEll}) performed above
for the snapshots in Figures~\ref{fig6},
in Figures~\ref{fig8} we plot the fitting parameters $a_\mathrm{fit}$,
$e_\mathrm{fit}$ and ${\phi_0}_\mathrm{fit}$, as a function of time, for
three data sets. Each data set spans a time window of
$200$~time-units at $\sub{t}{\rm end}=1.6\times10^6\, T_{\rm Prom}$
(blue data), $\sub{t}{\rm end}=3.2\times10^6\, T_{\rm Prom}$ (red data)
and $\sub{t}{\rm end}=6.0\times10^6\, T_{\rm Prom}$ (green data). The
fitted parameters display spurious short-time oscillations, whose
periodicity is related to the period of Prometheus in this
stroboschopic map ($200/(2\pi)\approx 32$). For $a_\mathrm{fit}$ and
$e_\mathrm{fit}$ we observe a longer time-scale variations, which
we attribute to secular effects. Note that,
while this secular time-scale is clear for $e_\mathrm{fit}$ in each
individual data set, for $a_\mathrm{fit}$ it appears only by comparing
the three data sets.

Regarding ${\phi_0}_\mathrm{fit}$, the slope of the linear trend is
an estimate of the apse-precession rate. We have fitted each data set to a
straight line (for the green data, this was done using the last
20 points). The average slope is an estimate for the
precession rate; the black lines displayed in Figure~\ref{fig8}(c)
have this average slope. We obtain
$\dot{\varpi}_\mathrm{fit}\approx 0.003867$~rad/time-units,
which corresponds to $2.27105$~deg/day. This estimate can be compared with
the nominal precession rate of the F ring which corresponds to
$\dot{\varpi}_\mathrm{Fring}=2.7001$~deg/day~\citep{BoshEtAl2002}.
The difference can be attributed to the simplicity of our model,
in particular the fact that we only considered the $J_2$ zonal harmonic.
The similarity in the slopes of the three data sets gives further
support to the claim that the ring displays apse alignment.

While our results show certain agreement with the observations,
we should emphasize that this
{\it a posteriori} consistency does not prove the dynamical
filtering that we have employed. It is through longer
numerical integrations, or a thorough understanding of the relevant invariant
phase space structures, that we can prove if the filtered particles indeed
escape and if the retained ones do not. With this proviso in mind, the
comparison with the observations is rewarding.

\subsection{Orbital resonances}

So far, we have shown numerically that, among the trapped particles (up to
$6.0\times 10^6 \, T_{\rm Prom}$), the most stable ones according to
the stability indicator $\Delta a$ yield a narrow and eccentric ring.
Here, we shall consider the orbital resonance structure in the region
where the ring is located ($2.321 < a < 2.3275$, c.f.
Figure~\ref{fig4}), considering all Saturn moons of our model, the
test particle and the $J_2$ zonal harmonic of Saturn.

We are interested in the semi-major axis $a$ of test particles that satisfy
a resonance condition~\citep{2002MorbidelliBook}, which can be written as
\begin{equation}
\label{resonances}
    \sum_{i=1}^4 \big[ k_i \dot{\lambda}_i - l_i\dot{\varpi}_i\big] =
    \sum_{i=1}^4 \big[ k_i f_i + (k_i-l_i) \dot{\varpi}_i\big] = 0.
\end{equation}
Here, all $k_i$ and $l_i$ are integer coefficients,
$\dot{\lambda}_i = f_i + \dot{\varpi_i}$
is the rate of change of the mean longitude,
$f_i$ and $\dot{\varpi}_i$ correspond to the mean motion
and the apsidal precession rate of Prometheus, Pandora, Titan and
the test particle, for $i=1,2,3,4$ respectively. We shall restrict
ourselves to consider orbital resonances with
$\sum_i |l_i| \leq 1$ (and $k_4\ne 0$) only, since all involved
eccentricities are quite small, and the amplitude of the Fourier
coefficient corresponding to the resonance is at least proportional to
$e_i^{|l_i|}$, with $e_i$ the eccentricity of the $i$-th
body~\citep{1999MurrayDermottBook}.

We locate the resonances as follows: We fix the frequencies for the moons
(from our numerical integrations, though similar results are obtained
using the nominal data) and use two fixed intervals, ${\cal I}_{f_4}$
and ${\cal I}_{\,\dot{\varpi}_4}$, for the possible frequencies of the
particle of the ring. These intervals contain the possible values of the
frequencies that are compatible with a semi-major axis $a_4$ in the interval
${\cal I}_{a_4}=[2.321,\ 2.3275]$; these intervals are given by
$f_4\in {\cal I}_{f_4} = [0.990365,0.994541]$ and
$\dot{\varpi}_4 \in {\cal I}_{\,\dot{\varpi}_4} = [4.470,4.498]\times 10^{-3}$.
Then, for a given set of integer coefficients $k_i$ and $l_i$ that
satisfy d'Alambert's relation, $\sum_{i}(k_i-l_i) = 0$~\citep{2002MorbidelliBook},
we check if zero is contained in the interval obtained from
Eq.~(\ref{resonances}). If zero is not in the interval, we conclude
that there is no resonance with the specific integer coefficients tested. If
the resulting interval does contain zero, we use Newton's method to compute
the test particle's semi-major axis that corresponds to the zero of
Eq.~(\ref{resonances}) together with the associated resonant frequencies. To
first order in $J_2$, the mean-motion and the horizontal epicyclic frequency
in terms of the semi-major axis are given
by~\citep{LissauerCuzzi1982}
\begin{eqnarray}
\label{freq_n}
n^2 & = & \frac{{\cal G} m_0}{a^3}
   \Big[ 1 + \frac{3}{2}J_2\Big(\frac{R_S}{a}\Big)^2\Big],\\
\label{freq_kappa}
\kappa^2 & = & \frac{{\cal G} m_0}{a^3}
   \Big[ 1 - \frac{3}{2}J_2\Big(\frac{R_S}{a}\Big)^2\Big].
\end{eqnarray}
From these frequencies, the apsidal precession rate is obtained,
$\dot{\varpi} = n-\kappa$.

\begin{figure}
  \centering
  \includegraphics[width=\linewidth]{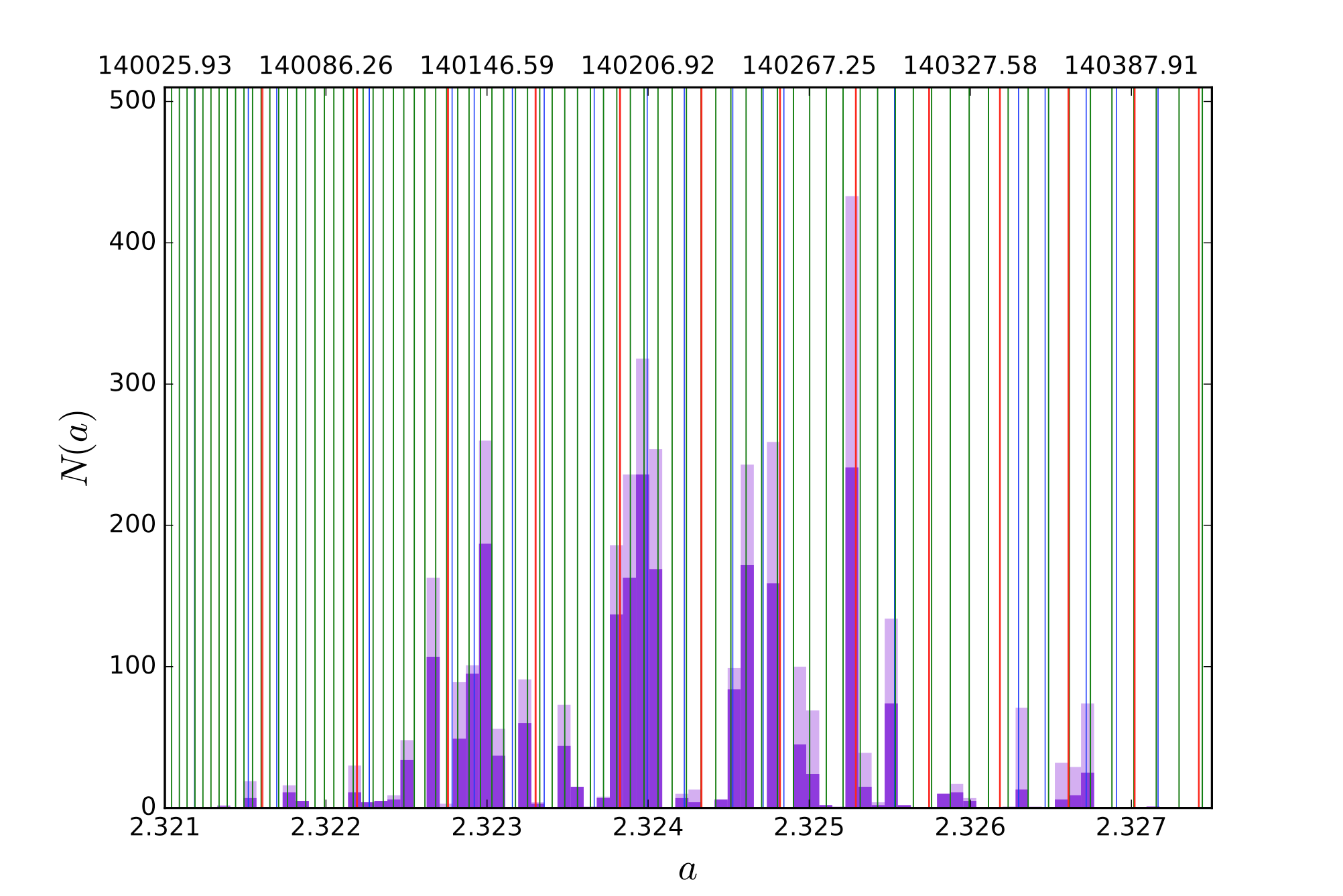}
  \caption{\label{fig9}
  Frequency histogram (using 80 bins) of the semi-major axis of the ring particles
  ($\Delta a < 10^{-5}$) for the the numerical integrations up to
  $\sub{t}{\rm end}=3.2\times10^6 \, T_{\rm Prom}$ (light purple histogram)
  and $\sub{t}{\rm end}=6.0\times10^6 \, T_{\rm Prom}$ (dark purple histogram).
  The vertical lines indicate the location of some resonances: green lines
  correspond to Prometheus' outer Lindblad and co-rotation resonances,
  red lines to  Pandora's inner Lindblad and co-rotation resonances,
  and blue lines to low-order orbital resonances involving the
  mean-motion of both shepherds and the ring-particles only.
  The lower horizontal axis scale is given in $R_S$
  and the upper one in~km.}
\end{figure}

In Figure~\ref{fig9} we show the histogram of the ring particle's
semi-major axis, using the numerical integrations displayed in
Figures~\ref{fig2} (light purple) and~\ref{fig4} (dark purple).
The histograms have been constructed using 80 bins in the semi-major
axis interval ${\cal I}_{a_4}$. The histograms show the decrease in the
ring particle number for the longer integrations mentioned above.

\begin{figure}
  \centering
  \includegraphics[width=0.9\linewidth]{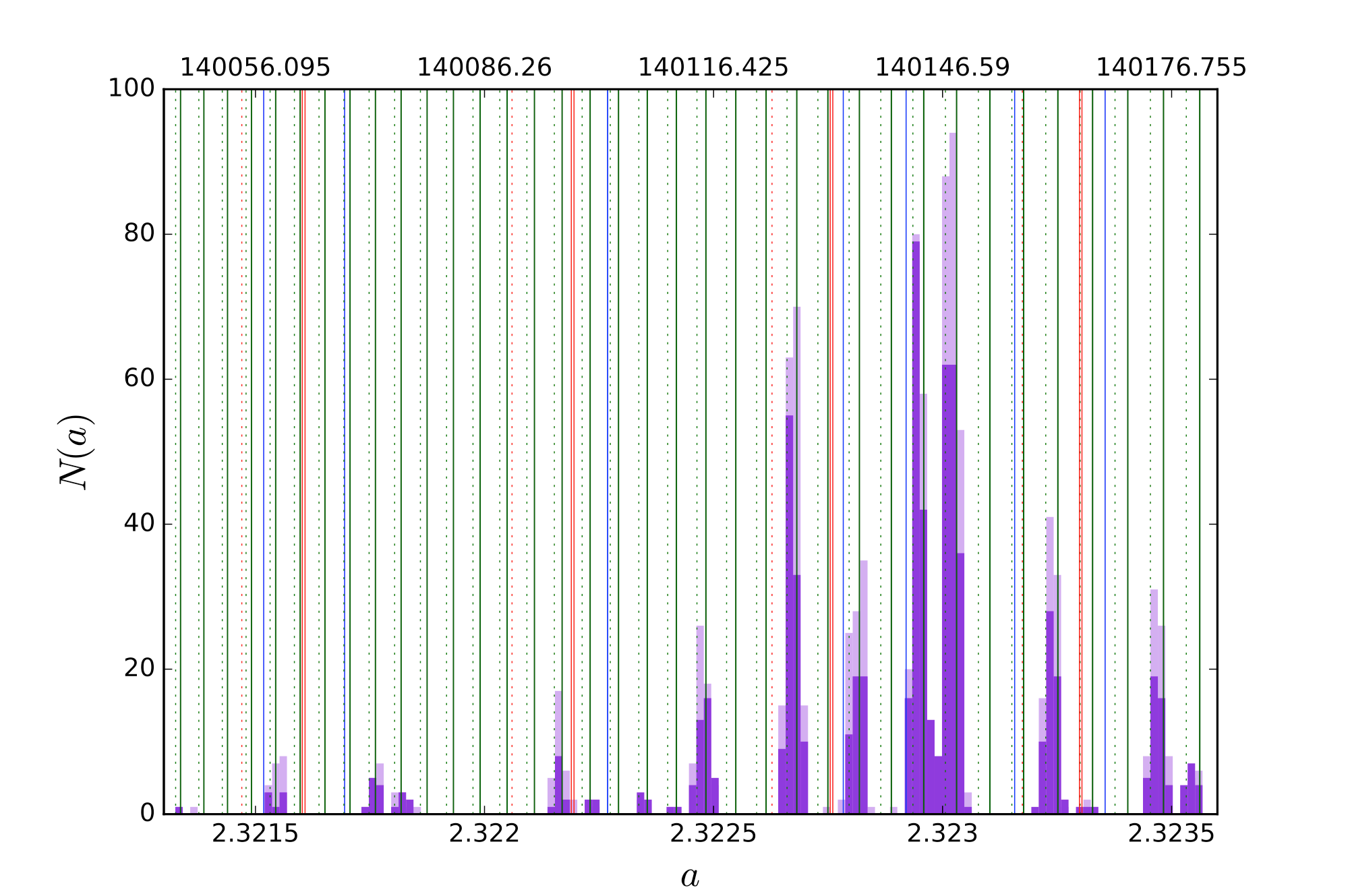}
  \includegraphics[width=0.9\linewidth]{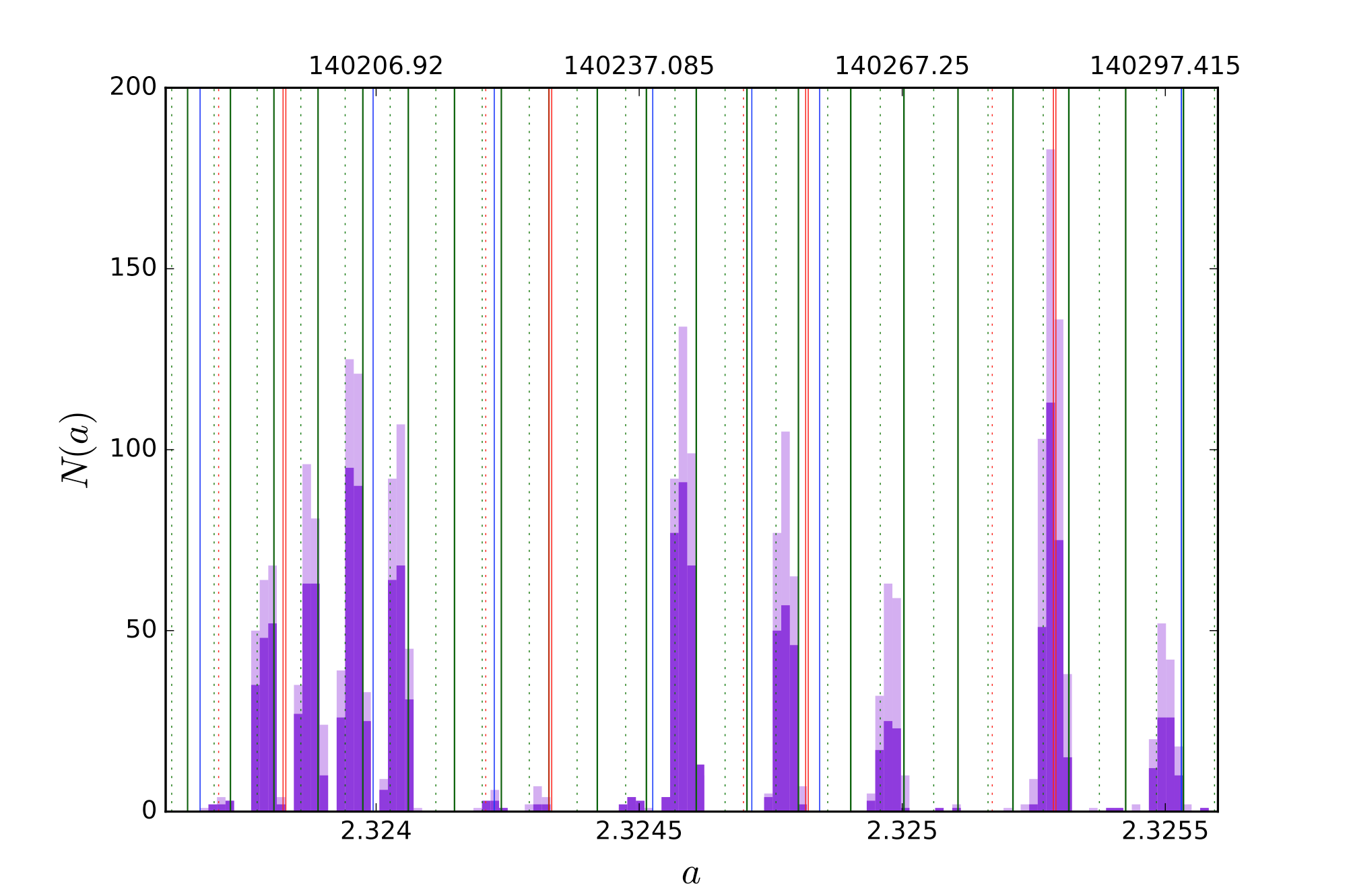}
  \caption{\label{fig10}
  Details of two regions of Figure~\ref{fig9}, using 400 bins to construct
  the histograms; note that the vertical scales differ. Some resonances
  are located at the gaps between two
  nearby regions of ring particle accumulation; most Pandora's inner
  Lindblad and co-rotation resonances (red lines) lie in these gaps. Some of
  Prometheus' outer Lindblad and co-rotation resonances (green lines), and
  resonances involving Prometheus' mean-motion and Titan's apse precession
  (dotted green lines), appear within the same region of accumulation
  of ring particles. The lower horizontal scales are given in $R_S$
  and the upper ones in~km.
  }
\end{figure}

The vertical lines displayed in Figure~\ref{fig9} correspond to different
resonances found within the semi-major interval of interest;
Figures~\ref{fig10} show the same results using 400 bins in the construction
of the histogram on ${\cal I}_{f_4}$; note that Figure~\ref{fig10}(b) spans the
region where the nominal semi-major axis of the F ring is located.
With the method described above, we located all low-order
resonances defined by the condition $\sum_i (|k_i|+|l_i|) \leq 40$.
The maximum order was set to 40 for convenience;
low order resonances are often preferred since the
amplitude of the Fourier coefficients decreases with the order of the
resonance. The blue vertical lines in Figures~\ref{fig9} and~\ref{fig10} are,
among these low-order resonances, those that only involve the mean
motions, i.e., $\sum_i |l_i| = 0$. It turns out that, up to the order
calculated, these resonances involve the mean motion of both shepherd
moons and the particle of the ring.

In addition to those resonances, the vertical green lines
correspond to Prometheus' outer Lindblad and co-rotation resonances,
$-j_1 f_1 + (j_1+1) f_4 + \dot{\varpi}_4 = 0$ and
$-j_1 f_1 + (j_1+1) f_4 + \dot{\varpi}_1 = 0$, respectively, with
$102 \leq j_1 \leq 179$. Likewise, the red lines correspond to
Pandora's inner Lindblad and co-rotation resonances,
$-(j_2+1) f_2 + j_2 f_4 - \dot{\varpi}_4 = 0$ and
$-(j_2+1) f_2 + j_2 f_4 - \dot{\varpi}_2 = 0$, respectively,
with $50 \leq j_2 \leq 62$. These pairs of resonances lie so
close together that it is difficult to distinguish them; only few
doublets of Pandora's resonances are apparent in Figures~\ref{fig10}.
During the computation
of these resonances, we noticed the proximity of other resonances
with nearby values of $j_1$ ($j_2$)
which involved the mean-motion of Prometheus (Pandora), the mean-motion
of the particle of the ring and the apse precession of Pandora (Prometheus),
or even Titan. While the former lie very close to Prometheus (Pandora)
outer (inner) Lindblad and co-rotation resonances, those that involve
Titan's precession rate are well separated.
In Figures~\ref{fig10} we have included the resonances of the form
$-j_1' f_1 + (j_1'+1) f_4 + \dot{\varpi}_3 = 0$ and
$-(j_2'+1) f_2 + j_2' f_4 - \dot{\varpi}_3 = 0$, that involve Titan's
precession rate,
as dotted-green and dotted-red vertical lines, respectively. The
latter are the only resonances displayed in Figures~\ref{fig9}
and~\ref{fig10} that involve Titan; some of them may be relevant
for the location of the ring, as we point out below. There are other
resonances in this region which also involve Titan; yet, those
resonances are three-body resonances, and their significance is not clear.

As shown in Figures~\ref{fig10}, many of the accumulation regions of the
ring particles can be associated with some resonances, and mostly
with Prometheus' outer Lindblad and co-rotation resonances (green lines).
We note that, to each such pair of resonances there is, within the
same region of ring-particle accumulation, a nearby resonance of the form
$-j_1' f_1 + (j_1'+1) f_4 + \dot{\varpi}_3 = 0$ which involves Titan's
apse precession. The most prominent peak, c.f. Figure~\ref{fig10}(b),
located at $a= 2.32529...\approx 140284.7$~km can be associated with
Pandora's inner Lindblad and co-rotation resonances with $j_2=57$; these
resonances are flanked by Prometheus' outer Lindblad and co-rotation
resonances with $j_1=119$ and their partner involving Titan's apse
precession ($j_1'=120$). Other Pandora's inner Lindblad and co-rotation
resonances (red lines) as well as most low-order three-body orbital
resonances (blue lines) appear in gaps between regions of accumulation
of ring particles. This is illustrated in Figure~\ref{fig10}(b)
by two of the gaps in the accumulation region close to the
semi-major axis $2.324$. One exception to this is the three-body resonance
$-11f_1-6f_2+17f_4=0$, which is located around
$a\sim 2.3263\approx 140345.7$~km; cf. Figure~\ref{fig9}.
The region close to $a=2.3255\approx 140297.4$~km seems to be
related to the proximity of the low-order three-body resonance
$-2 f_1 - f_2 + 3 f_4=0$ and Prometheus' outer Lindblad and
co-rotation resonances (with $j_1=117$), including the corresponding
partner resonance involving Titan's apse precession ($j_1'=118$).

From the observations described above, we may conclude that the
accumulation regions where the ring particles are found can be
essentially associated with Prometheus' Lindblad and co-rotation
resonances and their companion resonance which involves the precession
rate of Titan. Yet, not all Prometheus' outer Lindblad and co-rotation
resonances coincide with a region of accumulation of ring particles.
Similarly, Pandora's inner Lindblad and co-rotation resonances as well
as the low-order three-body resonances seem to be related to gaps rather
than regions of accumulation of test particles, but exceptions exist.
If these resonances somehow induce dynamical
instabilities, we may expect that as time passes the ring particles
found in the accumulation regions crossed by these resonances will
escape; this assertion may be proved by considering even longer
integrations. The explicit mechanism is not clear at this point.

\section{Summary and conclusions}
\label{Sec:Concl}

In this paper we have studied a simple model for the location of
Saturn's F ring. The model includes the gravitational influence of Saturn
with its $J_2$ zonal harmonic, Prometheus, Pandora and Titan, on
massless point particles, and assumes that the dynamics takes place in the
equatorial plane. The initial conditions of the test particles are
randomly chosen, imposing only that they are within the region between
Prometheus and Pandora at the initial time of our simulations. We have
computed accurate long-time numerical integrations that show that
there is a wide region, in the initial condition space, of test particles
that stay in the region between Prometheus and Pandora, i.e., that do not
escape from that region nor collide with the shepherds up to the final
integration time. Among these trapped test particles, we uncovered
a clear scale separation according to the value of the dynamical stability
indicator $\Delta a$. The most stable ones, which correspond to
$\Delta a < 10^{-5}$ with a maximum excursion $\lesssim 1.5$~km,
are located around
specific values of the semi-major axis, defining localized stripes in
the semi-major axis and eccentricity plane. Retaining only the most stable
test-particles, i.e., those particles whose dynamics do not display extended
radial excursions, their projection onto the $X-Y$ plane forms a narrow
eccentric ring which, as time passes, maintains a collective elliptic shape.

Fitting the whole ring to a Keplerian ellipse yields a semi-major axis that
is comparable to the observations. The fitted value for the
eccentricity is about an order of magnitude smaller, and the width
of the ring is $\sim 241$~km, comparable to some observations and
simulations, but largely overestimating a core of $1-40$~km. Using
various snapshots spanning $\sim 30$ periods of Prometheus at different
epochs, we obtain consistent values for the apse precession rate,
which are somewhat smaller than the observations, but of the same
order of magnitude. In that sense, the ring obtained displays
apse alignment. We emphasize that, in the context of our
simulations, it cannot be attributed to the mass of the ring;
the actual mechanism that yields apse alignment remains unclear.

Note that the actual value for the semi-major axis obtained
from fitting the rings of Figures~\ref{fig6} to a Keplerian ellipse,
$a_{\rm fit}\approx 2.3241\,R_S$, does not match the location
of any of the (most populated) ring-particle accumulation regions;
c.f. Figure~\ref{fig10}. This remark is trivial, in the sense that
the fitting procedure is collective and involves the position of
all ring particles with equal weight. Yet, it emphasizes the fact
that the whole set of stable trapped ring particles has to be
considered. This indicates also that orbital fits of individual
objects (which a priori are located within a ring-particle accumulation
region) may not match the fitting parameters for the ring. This
agrees with results of~\citet{Cooper2013} that show that local
variability of the ring is compatible with a consistent solution
for the whole ring. It also suggests that the orbital elements fitted
to some bright features that have been observed, represent a
region where stable ring-particles accumulate.

\citet{CuzziEtAl2014} addressed similar questions considering a
more realistic model (14 Saturn moons and including $J_2$, $J_4$ and
$J_6$ Saturn's zonal harmonics) and shorter time integrations, up to
$20000$ Prometheus periods. As a measure of the stability of test particles
they considered the semi-major axis RMS deviation during the time
spanned by the integration. In their Figure~9, they obtain a one to one
correspondence of the trapping stripes with the Prometheus' outer
Lindblad and co-rotation resonances. (The aim of that
figure was to show the robustness of the calculations with respect
to changes of the zonal harmonics.) While we also find an
association of the regions of ring-particle accumulation to such
Prometheus resonances, our results show that not all Prometheus
resonances are associated with a ring-particle accumulation region.
In this case, the difference is attributed to the facts that
Pandora was not included in those calculations and the integrations
are rather short (5000 Prometheus periods for their Fig.~9). Other
calculations that include Pandora
\citep[][Figures~14--16, up to $20000\,T_P$]{CuzziEtAl2014}
show that Pandora's resonances influence the stability of the tests
particles, rendering them less stable, i.e., larger semi-major axis
RMS deviations are observed. Our results are similar with respect to
the semi-major axis, though $\Delta a$ defines naturally the stability
through a scale separation. We also observe more sensitivity with
respect to the eccentricity of the ring particles, which in our case
are smaller than the values considered in \citeauthor{CuzziEtAl2014}
calculations; this may be related to the much longer integration times
we considered. Despite of the specific differences of the models and
methods, the results are consistent. We interpret this as the robustness
of the approach.

The quantity $\Delta a$ seems to be a good dynamic indicator to address if a
test particle escapes or not, depending on whether or not its value is large
enough. In our numerical results, there is a natural scale separation defined
by the statistics of $\Delta a$ for the trapped particles. However, some
ring-particles with a small value of $\Delta a$ may escape at later times.
In our model, escape is the result of the accumulation of small radial
excursions due to the combined action of the moons considered
in the model, which are enhanced by impulsive encounters due to the
proximity of the shepherd moons.

The ring particles accumulate in regions which seem related to
occurrence of orbital resonances, typically involving Prometheus'
outer Lindblad and co-rotation resonances, though not all
such resonances are associated with regions of accumulation; similar
results were obtained by \citet{CuzziEtAl2014}. We also considered
the location of other resonances. Low-order resonances in the region
of the ring involve the mean motions of both shepherds and the particle
of the ring; these resonances, as well as most of Pandora's inner Lindblad
and co-rotation resonances, seem related to gaps where the radial
excursions are large enough, though some match ring-particle accumulation
regions.

We emphasize that we have only found a possible association between
resonance locations and the occurrence of accumulation regions and gaps,
without providing an explanation that links the specific resonance
to the actual dynamics. \citet{CuzziEtAl2014} propose an
``anti-resonance'' mechanism to explain the suppression of radial
diffusion. In essence, the idea is that at certain locations
($m$-th Prometheus' outer Lindblad and co-rotation resonances,
with $m\gg 1$) radial excursions promoted by perturbations from a
close approach of Prometheus to the ring, are counterbalanced
by the next close approach of this shepherd. \citet{CuzziEtAl2014}
remark that the condition to suppress radial diffusion within the
antiresonance mechanism is equivalent to the empirical commensurability
$f_{\rm Prom}-f_{\rm Fring} \approx 2\dot{\varpi}_{\rm Fring}$ noted
by~\citet{Cooper2013}. While this mechanism is physically appealing, the
fact that it is equivalent to the empirical commensurability which
{\it does not} satisfy d'Alambert's relation, suggests that something
is still missing. In our analysis we have observed that
close to Prometheus' outer Lindblad and co-rotation resonances,
resonances involving the same (Prometheus and the ring particle)
mean-motions but that the precession rate of Pandora or Titan,
in particular the latter, seem to be relevant to define the narrow
regions where the ring particles accumulate, though the actual mechanism
is not clear. It is tempting to consider these resonances to complete
the anti-resonance mechanism; this is left for the future.

Our model is rather simple and does not include many important
contributions for a realistic comparison with the F ring. The role of
Titan, as displayed by the resonance structure, seems marginal; yet,
it influences the actual location and order of Prometheus'
resonances~\citep{CuzziEtAl2014}, and some resonances involve,
which somehow makes it relevant. While Titan promotes
radial excursions, it seems unable to clear all trapped unstable
particles in a short-time scale; in particular, the region beyond
the location of the F ring. Interactions with other moons
or ring-particle collisions may contribute to this aspect. A
point to be emphasized is that the occurrence of narrow rings
does not require very specific scenarios with regards to the
parameters, but certain stability conditions.

\section*{Acknowledgments}
We would like to acknowledge financial support provided by the projects
IG--100616 (PAPIIT, UNAM), SC15-1-IR-61-MZ0 and SC16-1-IR-54 (DGTIC, UNAM),
the spanish grant
MTM2015-67724-P and the catalan grant 2014 SGR 1145.
LB expresses his gratitude to the Marcos Moshinsky Foundation for the
financial support through a ``C\'atedra de Investigaci\'on 2012''. It is our
pleasure to thank Carles Sim\'o for his encouragement, valuable comments
and discussions. We would also like to thank the careful review, criticism
and comments from Carl D. Murray.

\begin{figure*}
  \centering
  \includegraphics[width=0.65\linewidth]{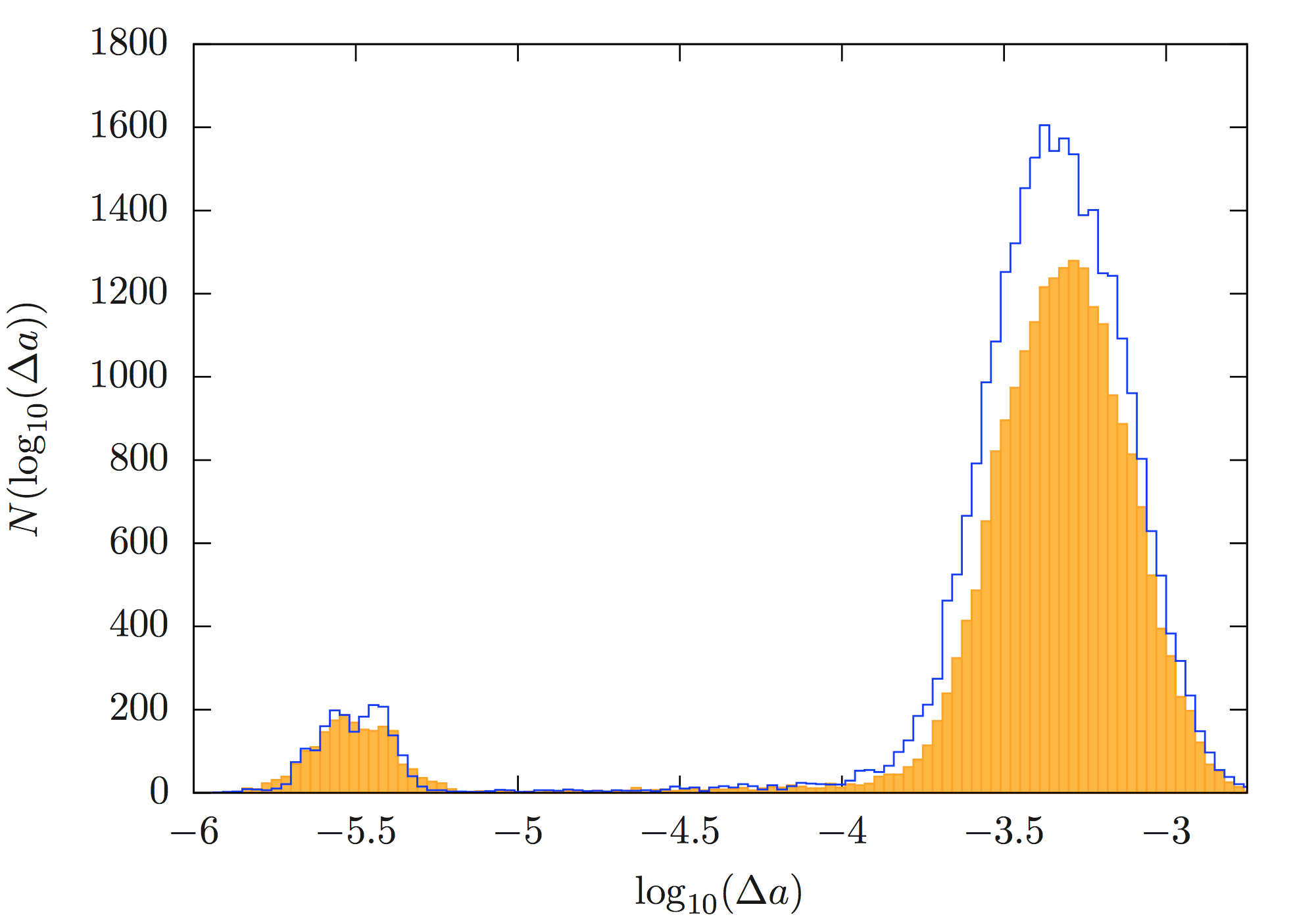}
  \includegraphics[width=0.65\linewidth]{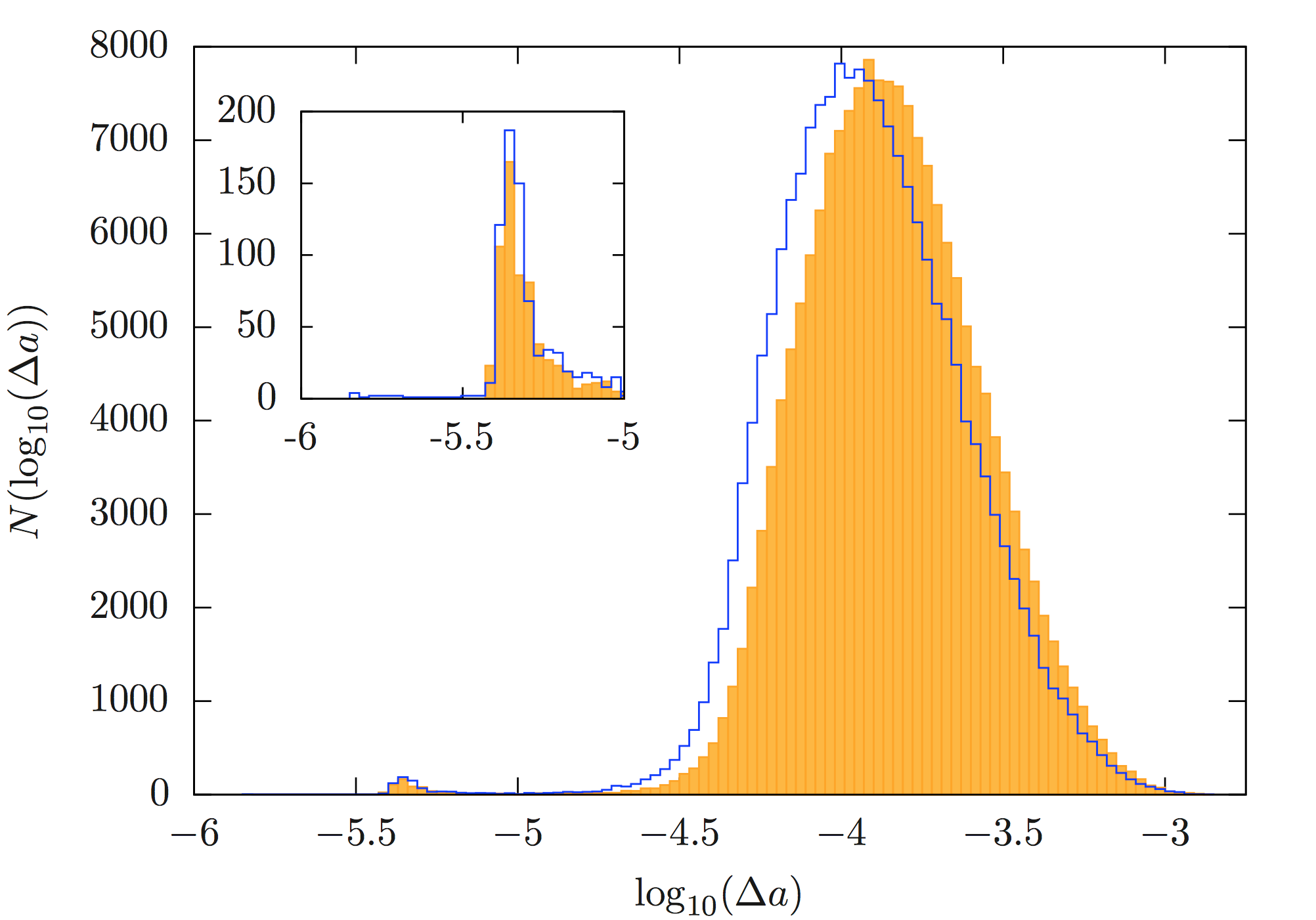}
  \caption{\label{fig11}
  Frequency histogram of $\log_{10}(\Delta a)$ for two regions
  of the semi-major axis for integrations up to
  $\sub{t}{\rm end}=1.6\times 10^6 \, T_{\rm Prom}$. Filled
  histograms correspond to simulations that include Titan,
  the empty ones are without it. (a) Semi-major axis region
  around $a\sim 2.32475\approx 140252.2$~km, which
  contributes to the ring. (b) Semi-major axis region around
  $a\sim 2.33135\approx 140650.3$~km; the inset shows an
  enlargement for test particles satisfying the stability criterion.
  Note that the vertical scales are different.
  }
\end{figure*}

\appendix
\section{Numerical simulations with and without Titan}
\label{Sec:appendix}

Here, we present some numerical results to justify the introduction
of Titan and show that it promotes radial excursions for those
test particles not located within the ring-particle accumulation stripes.
To this end, we consider similar integrations as described before,
including or not the gravitational influence of Titan. The time integrations
extend up to $\sub{t}{\rm end}=1.6\times 10^6\,T_{\rm Prom}$, and
the initial conditions are set in two small semi-major axis regions
spanning $\delta a\approx 66.4$~km; the remaining test-particle orbital
elements are set as before. The first region
($a\sim 2.32475\approx 140252.2$~km)
contains stable stripes where ring particles accumulate, i.e., some
particles belonging to the ring of Figures~\ref{fig6}. The second
region covers larger values of the semi-major axis
($a\sim 2.33135\approx 140650.3$~km) which displays trapped
particles, with few of them being stable according to our
criterion ($\Delta a< 10^{-5}$). Note that $\sub{t}{\rm end}$ in these
integrations is shorter than those considered in Figures~\ref{fig2}
and~\ref{fig4}; we thus expect that the particles in the second region
satisfying the stability criterion will eventually display some
instabilities.

In Figures~\ref{fig11}, we present the frequency histograms
of $\log_{10}(\Delta a)$ corresponding to the two intervals
of semi-major axis described above; the histograms of simulations
that include Titan correspond to the filled ones, while those
without it are the empty ones. Figure~\ref{fig11}(a) illustrates
the results for the first region defined above. Noticeably, there
are more particles trapped when Titan is not included in the
simulations, though about the same number of particles that fulfill
the stability criterion. In turn, Figure~\ref{fig11}(b) displays
the results for the semi-major axis region around
$a\sim 2.33135\approx 140650.3$~km. In this case, we find about the
same number of trapped and ring-particles, seemingly independently
of Titan's presence, though marginally more for the simulations that
do not include it. Notice that for the trapped particles with
$\Delta a>10^5$, the histograms without Titan (empty histograms)
are slightly shifted to the left. That is, trapped particles are
more stable in the simulations that do not include Titan.

These results show that Titan indeed promotes radial excursions for
particles not satisfying our stability criterion.
For semi-major axis in the region where the accumulation stripes are
found, Titan's inclusion allows a more efficient escape mechanism.
Moreover, the ring-particle accumulation regions are slightly
shifted, as noted by \citet{CuzziEtAl2014}.
For larger values of the semi-major axis, the presence of Titan
induces instabilities, though not enough strong to swipe
the trapped particles away.

\section*{References}
\bibliographystyle{elsarticle-harv}
\biboptions{authoryear}
\bibliography{bib_rings}

\end{document}